\documentclass[%
 reprint,
 amsmath,amssymb,
 aps,
floatfix,
prd]{revtex4-2}

\usepackage[utf8]{inputenc}
\usepackage[english]{babel}
\usepackage{dcolumn}
\usepackage[dvipsnames]{xcolor}
\usepackage{bm}
\usepackage[T1]{fontenc}

\usepackage{graphicx}
\usepackage{verbatim}
\usepackage{cprotect}

\usepackage{amsmath}
\usepackage{textcomp}
\usepackage{gensymb}
\usepackage{braket}
\usepackage{amsthm}
\usepackage{amssymb}
\usepackage{mathrsfs}
\usepackage[normalem]{ulem}

\numberwithin{equation}{section}

\usepackage{hyperref}

\newcommand{\f}{\frac}
\renewcommand{\d}{\text{d}}

\begin{document}

\preprint{APS/123-QED}

\title{Frequency shift and viewing direction variations in
gravitational lensing}

\author{Mikołaj Korzyński}
 \email{korzynski@cft.edu.pl}
\affiliation{Center for Theoretical Physics, Polish Academy of Sciences \\ Al.\,Lotników 32/46, 02-668 Warsaw, Poland}

\author{Mateusz Kulejewski}
\email{m.kulejewski@uw.edu.pl}
\affiliation{Faculty of Physics, University of Warsaw \\ Pasteura 5, 02-093 Warsaw, Poland}



\date{\today}

\begin{abstract}
In a gravitational lensing system, the relative transverse velocities of the lens, source, and observer induce a frequency shift in the observed radiation. While this shift is typically negligible in most astrophysical contexts, strategies for its detection have been proposed for both electromagnetic and gravitational waves. This paper provides a rigorous theoretical treatment of the effect, deriving general expressions for the frequency shift within a lensing system embedded in a cosmological spacetime. Our formulation remains valid for arbitrary distances and velocities—including highly relativistic regimes—under any Friedmann-Lemaître-Robertson-Walker metric.

Expanding upon previous papers on moving lenses, we provide a detailed analysis of frequency effects induced by lenses moving at relativistic speeds. Furthermore, we extend standard lensing theory by deriving an exact formula for the variation in the source's viewing direction. This result is of interest for strongly anisotropic emitters, such as compact binary systems emitting gravitational waves. Finally, we quantify the apparent misalignment between the lens and the source's two images produced by time-delay effects in lens systems moving with high velocity. 
\end{abstract}

\maketitle

\tableofcontents

\renewcommand{\S}{\mathcal{S}}
\renewcommand{\O}{\mathcal{O}}
\newcommand{\LL}{\mathcal{L}}
\newcommand{\E}{\mathcal{E}}
\newcommand{\PP}{\mathcal{P}}
\newcommand{\QQ}{\mathcal{Q}}
\newcommand{\p}[2]{\frac{\partial #1}{\partial #2}}

\newcommand{\beq}{\begin{equation}}
\newcommand{\eeq}{\end{equation}}
\newcommand{\bea}{\begin{eqnarray}}
\newcommand{\eea}{\end{eqnarray}}
\newcommand{\bean}{\begin{eqnarray*}}
\newcommand{\eean}{\end{eqnarray*}}

\newcommand{\aap}{Astron. Astrophys.}                
\newcommand{\apjl}{Astrophys. J. Lett.}              
\newcommand{\apjs}{Astrophys. J. Suppl. Ser.}        
\newcommand{\mnras}{Mon. Not. R. Astron. Soc.}       
\newcommand{\aj}{Astron. J.}                         
\newcommand{\araa}{Annu. Rev. Astron. Astrophys.}    
\newcommand{\pasp}{Publ. Astron. Soc. Pac.}          

\newcommand{\const}{\textrm{const}}

\newcommand{\MK}[1]{{\color{MidnightBlue}{[{\bf MK}: #1]}}}
\newcommand{\edMKK}[1]{{\color{BrickRed}{{#1}}}}
\newcommand{\edMK}[1]{{{#1}}}

\newcommand{\TODO}[1]{{\color{PineGreen}{{\bf TO DO:} #1}}}
\newcommand{\highlight}[1]{#1}

\newcommand{\gvec}[1]{\vec{\hspace {0.2mm} #1} \hspace{0.2mm}}

\def\mathdefault#1{#1}
\everymath=\expandafter{\the\everymath\displaystyle}
\newif\iffigures
 \figurestrue   

\newif\iffancyfigures
\fancyfiguresfalse

\section{Introduction}

Gravitational lensing is caused by the deflection of a light ray by the gravitational field of a massive body. In a static lensing setup  photon energy is conserved during the interaction with the  gravitational field of the lens, which results in the radiation frequency remaining unchanged by the deflection.   However, if the source, the observer or the lens are in relative motion, we may expect the lensing to induce a change in the wave frequency as well. Under the geometrical optics approximation, which is valid most of the time in astrophysical settings, the effect is identical for electromagnetic and gravitational radiation. 

This frequency shift, or the spectroscopic effect of lensing,  has recently attracted significant interest from the astrophysical community,  particularly in the context of gravitational wave lensing. It turns out that analyzing the interference effect of signals corresponding to two or more strongly lensed images offers a promising way to measure the transverse velocities of the source and the lens \cite{SuyamprakasamBejgeretal, PhysRevD.109.024064, Itoh.PhysRevD.80.044009, Samsing:2024xlo, Ubach:2025oxr}, although it requires the identification of the signal first \cite{Chan:2024qmb, Chan_2026}.
In the electromagnetic domain, it has been proven that microlensing by moving lenses can cause spectral line displacements and profile changes whose magnitude is within the reach for the next generation high precision spectrographs. It has also been suggested that the measurement of this effect can provide additional information about the lens and break the degeneracy between the lens parameters in microlensing events \cite{PhysRevD.101.024015}, complementing  other techniques such as the variations of the apparent position of the source (astrometric microlensing) \cite{1995A&A...294..287H, 1995ApJ...453...37W, miyamoto1995astrometry, 2000ApJ...534..213D, 2017Sci...356.1046S}, interferometry \cite{refId0, Dalal:2002ki, wyrzykowski1} or the microlensing parallax effect \cite{10.1093/mnras/134.3.315, Gould1994_Parallax, 1995ApJ...454L.125A, 2015ApJ...799..237U}.

The study of the theory of the frequency shift in gravitational lensing has a long history dating back to Pyne and Birkinshaw \cite{1993ApJ...415..459P}. The same effect was also considered by Bertotti and Gampieri \cite{B_Bertotti_1992} in a different setting, as a correction to the Doppler frequency shift of a signal emitted by a spacecraft located in the Solar System. 
Through the late 1990s and early 2000s, the topic was studied mainly in conjunction with the problem of computing the bending angle of a moving lens. It was known that the lens motion affects the deflection angle, but the formula of the velocity correction obtained by Capozziello \textit{et al.} \cite{CAPOZZIELLO199911} was inconsistent with the result of Fritelli, Kling and Newman \cite{PhysRevD.65.123007}. The discrepancy was later resolved in \cite{fritelli2003, fritelli2003-2}. Paper \cite{fritelli2003} also contains the derivation of the frequency shift under the assumption of small lens velocity and a flat background metric. Concurrently, Kopeikin and Sch\"affer developed a Lorentz-invariant theory of light propagation in the gravitational field of moving point masses in the post-Minkowski approximation independently, deriving an expression for the frequency shift effect \cite{PhysRevD.60.124002}.
Later, Wucknitz and Sperhake derived the deflection angle for a lens with arbitrary velocity \cite{PhysRevD.69.063001} and the equation for the variation of frequency for small velocities.

In most prior work,  the frequency effect is computed using the first-order-in-velocity approximation (FOV) in the terminology of Wucknitz and Sperhake \cite{PhysRevD.69.063001}, valid for velocities much smaller than the speed of light. The only  exception we are aware of is  \cite{PhysRevD.60.124002}, which uses arbitrary velocities of the source, observer and lens. However, the background is assumed to be flat Minkowski spacetime rather than a more general cosmological solution, and the lens model is limited to  a collection of point masses. 

In this paper, we aim to derive the most general formula for the frequency effect of gravitational lensing with arbitrary motion of the  source, observer and lens. Our derivation is based on fairly general and physically motivated assumptions:
\begin{itemize}
    \item The entire lensing system is embedded in a Friedman-Lemaitre-Robinson-Walker metric describing the large-scale geometry of the Universe.
    \item We assume the thin lens approximation and the small deviation angle approximation (first-order-in-deviation approximation, FOD). We make use of the standard definitions of the lensed and unlensed angular position of the source as well as the deflection angle.
    \item We assume that the metric in the vicinity of the lens is static \emph{in the lens rest frame}, and therefore, the frequency of the deflected radiation is strictly conserved \emph{in the lens rest frame}.
\end{itemize} 
Crucially, we do not assume the motions of the lens, source or observer to be slow with respect to the speed of light, nor do we  make any restricting assumptions regarding the geometry of the lens or its lensing potential.

 Another significant novel component of this work is the rigorous computation of the variation of the viewing direction of the source due to lensing. This variation is absent from the standard gravitational lensing formalism as described in the literature \cite{1992grle.book.....S, PettersLevineWambsganss, 2006glsw.conf.....M, Blandford_Narayan_1986}. The main reason is that this variation is fairly small and its effect is negligible for electromagnetic sources, whose emission tends to be fairly isotropic. However, it is crucial for the frequency variation effect.  Moreover, it may also be important for highly anisotropic sources, such as binary systems emitting gravitational waves. For this type of GW sources, a change in the viewing direction corresponds directly to a variation in the phase and polarisation of the wave. The viewing direction formula we derive extends the standard lensing theory in an FLRW spacetime, predicting additional phase differences between multiple lensed images that may become observable by next-generation GW detectors. This general formula constitutes one of the key original results of this paper.

This paper significantly generalizes previous findings regarding the frequency variations in gravitational lensing systems featuring moving lenses, sources and observers. In particular, our first aim is to extend the results of the  aper by Wucknitz and Sperhake by lifting the FOV approximation from Sec. IVC of \cite{PhysRevD.69.063001} and thus allowing the velocities to be arbitrary. We also explicitly include the effects of the transverse displacement of the source from the line of sight, represented by the $\beta^{\bm A}$ term in Eq. \eqref{eq:DZ_from_angles5}. This term is particularly important for sources with large angular extension, in which the frequency shift effect may vary across each image due to the presence of this term. It also plays an important role in microlensing, in which the relative position of the source and the lens changes appreciably over the observation time. Thus, even though the angular size of the source is small, including the $\beta^{\bm A}$ term is necessary to compute the time dependence of the observed frequencies of the two images.  

Our work also extends the results of Kopeikin and Sch\"affer \cite{PhysRevD.69.063001}, who calculated, among other things, the spectroscopic effect of lensing in the post-Minkowski approximation. We improve on this by embedding the lensing setup in a more realistic cosmological FLRW solution rather than a static Minkowski spacetime, incorporating cosmological effects. We also consider a more general type of deflectors, allowing for an arbitrary mass distribution, rather than a collection of point masses. In comparison to studies like \cite{PhysRevD.101.024015}, which focus on microlensing of a distant source, our formalism allows the source and the lens to be located at any distances from the observer and automatically includes the effects of the FLRW metric if the distances in question are sufficiently large. The frequency shift formula in \cite{PhysRevD.101.024015} is also derived under the FOV approximation, unlike this paper. Moreover, to our knowledge, this is the first work  considering the variations of the viewing direction in a lensing system in full generality.

Our derivations rest on two pillars.  The first one is the standard gravitational lensing theory, where the two-dimensional deflection angle is treated as an arbitrary function of the two-dimensional lensed position of the source. The second one is  the propagation of light between the source and the lens and between the lens and the observer, which is modeled here using the first order geodesic deviation equation (GDE) or, more specifically, using its resolvent, known the bilocal geodesic operator (BGO)  \cite{Korzynski:2017nas, Korzynski:2024jqt}, Jacobi propagator \cite{Vines:2014oba, Dixon2, DeWittBrehme, Flanagan:2018yzh, Flanagan:2019ezo, PhysRevD.83.083007}, or the Green's function for the GDE \cite{1993ApJ...415..459P}. An important guiding principle is that at each stage we seek expressions that are explicitly Lorentz- (or frame-) invariant. This approach naturally ensures that the results are valid beyond the FOV approximation in all velocities involved. \edMK{We build here on our previous results about the application of the GDE around a null geodesic to geometric optics and the geometry of null frames associated with it \cite{Grasso:2018mei, Korzynski:2024jqt, PhysRevD.101.063506}.}

Our derivation shares some ideas with the approach of Pyne and Birkinshaw \cite{1993ApJ...415..459P}. The similarity lies in modeling the propagation of light rays between the source and the lens and between the lens and the observer using the geodesic deviation equation and making use of its resolvent. Again, the crucial difference is that we allow the ambient spacetime to be a cosmological solution rather than a flat spacetime. Furthermore, our method of derivation concerning the variation of the viewing direction bears conceptual similarity to the approach presented in Sec. IVC  of \cite{PhysRevD.69.063001}.

The most important results of this paper are as follows:
\begin{itemize}
    \item the Eq. \eqref{eq:DZ_from_angles5} for the redshift variation in terms of the  standard angles as defined in the lensing theory, together with the definition of the covariant transverse velocity \eqref{eq:Delta_u_def}  and \eqref{eq:Sigmadef}, as well as its equivalent forms \eqref{eq:DZ_from_angles4} and \eqref{eq:DZ_from_thetaalpha_UO}
    \item the equation for the variation of the viewing direction  \eqref{eq:iota_from_theta_beta}, together with \eqref{eq:nu_def}.
\end{itemize}

The paper is organized as follows: we end this section with a short review of the notation and conventions used in this paper. Section \ref{sec:geometric_optics_in_FLRW_spacetimes} reviews the fundamental concepts and results in geometric optics and geometry of null geodesics in FLRW spacetimes. This includes the theory of the geodesic deviation equation and the bilocal geodesic operators (BGO). We define the lensing setup in the FLRW spacetime, derive the standard lensing equation and present the first technical result:  formula \eqref{eq:DZ_from_angles3} for the redshift variation in terms of the three key angles: the lensed position, the deflection angle and the variation of the viewing direction. In Sec. \ref{sec:variation_of_the_viewing_direction}, we utilize the BGO framework to derive the general formula for the variation of the viewing direction.

The following Sec. \ref{sec:spectroscopic_effect_of_lensing} constitutes the central part of this paper, where we combine the results of Secs. \ref{sec:variation_of_the_viewing_direction} and \ref{sec:geometric_optics_in_FLRW_spacetimes} to obtain the general formula for the frequency shift. Next, we discuss the physical consequences of the formula and the dependence of the frequency shift on the relative motions of the lens, source, and observer. We also define the fully relativistic counterpart of the Doppler velocity from \cite{Kayser1986, PhysRevD.69.063001} and  estimate the frequency difference between multiple images of a point source. We then apply the results to a relativistic lens (massive object moving with velocity close to the speed of light) and estimate the spectroscopic effect of lensing by such a body.

In Sec. \ref{sec:flat_spacetime} , we consider the important case of a flat spacetime, relevant for microlensing. We also discuss the effect of misalignment of the images and the lens for a spherical lens in motion.
We conclude the paper with a summary in Sec. \ref{sec:summary}. The derivations of some of the more technical results are contained in the Appendixes.

\subsection{Notation and conventions}

We assume that the speed of light $c=1$, unless explicitly stated otherwise. We work in the signature $(-,+,+,+)$. 
Bilocal geometric operators (BGOs), which give the solution of the geodesic deviation equation at $\lambda_1$ given the initial data at $\lambda_2$, see Eqs. \eqref{eq:WXX_WXL} and \eqref{eq:WLX_WLL}, are denoted by $W_{**}(\lambda_1, \lambda_2)$ or, equivalently, $W_{**}^{1\leftarrow 2}$, with the indices $**$ taking the values $XX$, $XL$, $LX$ or $LL$. 

\emph{The fiducial geodesic. } the notation $\tilde{l}$ refers to the tangent vector to the fiducial null geodesic, while $l$ refers to the  geodesic perturbed by the lens, $l = \tilde{l} + \Delta l$.

\emph{Indices.} We use two types of indices in this paper: those which correspond to a coordinate system, and thus a tetrad, or basis, related to those coordinates, and indices related to a special tetrad unrelated to the coordinates, but rather aligned with a light ray. The Latin boldface indices $\bm{A}, \bm{B}$, etc., refer to the two transverse tetrad components $1,2$, while the Greek indices $\mu, \nu$, etc., refer to all four components $0, 1, 2, 3$. We also use the notation $X^l$ for the component directed along the null geodesic,
i.e. $X = X^{\bm A}\,e_{\bm A} + X^l\,\tilde l$, with $e_{\bm A}$ being the transverse vectors. The dot product $\cdot$ refers to the spacetime product $X \cdot Y \equiv X^\mu Y_\mu \equiv g_{\mu\nu}X^\mu Y^\nu$.

\section{Geometric optics in FLRW spacetimes}\label{sec:geometric_optics_in_FLRW_spacetimes}

\subsection{Tetrads, geodesic deviation equation and bilocal geodesic operators} \label{sec:tetrads}

The propagation of light between the source and the lens and between the lens and the observer takes place in an FLRW metric. Since the size of the lens and the source are small with respect to the size of the Universe, we can approximate the behaviour of the light rays in this region by the first-order geodesic deviation equation (GDE). This way, we are able to take advantage of its linearity and thus the linear dependence of the perturbed geodesic on the initial data.

 The general solution of the GDE for null geodesics can be expressed easily in terms of the bilocal geodesic operators (BGO), also known in the literature as the transition matrix or Green's function  or Jacobi propagators.  They connect the variation of a geodesic at a given point to its variation at all other points. Below we now review shortly the basic results of the theory of the GDE and the BGOs we use in this paper. For a longer discussion of the formalism and detailed derivations we recommend \cite{Grasso:2018mei, Korzynski:2024jqt}.

The GDE is an equation for a vector field $\xi^\mu$, defined along a fiducial null geodesic $\gamma_0$,
\beq
\nabla_l \nabla_l \xi^\mu - R^\mu{}_{\nu\alpha\beta}\,l^\nu\,l^\alpha\,\xi^\beta = 0, \label{eq:GDE}
\eeq
where $l^\mu$ is the null tangent and $\lambda$ is the corresponding affine parametrization of the geodesic. \edMK{Assume we perturb the geodesic at the observation point $\O$ via $x^\mu_\O \to x^\mu_\O + \delta x^\mu_\O$, $l_\O^\nu \to l_\O^\mu + \delta l_\O^\mu$. We introduce the covariant direction deviation via $\Delta l_\O^\nu = \delta l_\O^\nu + \Gamma^\mu{}_{\alpha\beta}(\O)\,l_\O^\alpha\,\delta x_\O^\beta$. The reader may verify that $\Delta l_\O^\mu = \nabla_l \xi^\mu(\cal O)$, i. e. the covariant variation of the tangent vector at the linear order is exactly equal to the covariant derivative of the corresponding perturbation vector \cite{Grasso:2018mei, Korzynski:2024jqt}. With this, the deviation vector $\xi^\mu$ at any different point can be computed by solving the GDE with the initial data of the form
\bea
\xi^\mu(0) &=& \delta x_\O^\mu \\
\nabla_l \xi^\mu(0) &=& \Delta l_\O^\mu.
\eea
At any other point $\lambda_1$ the perturbation has the form of  $\delta x^\mu = \xi^\mu(\lambda_1)$, and the perturbation of the tangent vector is $\Delta l^\mu = \nabla_l \xi^\mu(\lambda_1)$.

The relation between the initial data and the solution of the GDE is strictly linear, because the GDE is linear. Therefore, we can also write
\bea
\label{eq:WXX_WXL}
\delta x(\lambda) &=& W_{XX}(\lambda,\lambda_\O)\,\delta x_\O + W_{XL}(\lambda, \lambda_\O)\,\Delta l_\O, \quad \quad
\\
\label{eq:WLX_WLL} 
\Delta l(\lambda) &=& W_{LX}(\lambda,\lambda_\O)\,\delta x_\O + W_{LL}(\lambda, \lambda_\O)\,\Delta l_\O, \quad \quad
\eea
with four linear operators $W_{XX}$, $W_{XL}$, $W_{LX}$, $W_{LL}$, we call the bilocal geodesic operators (BGOs). These operators in turn can be expressed as solutions of appropriate ODEs involving the curvature \cite{Korzynski:2024jqt}, but this does not play a role here. Together they form the resolvent of the geodesic deviation equation. The subscripts in the BGOs describe their functions in Eqs. \eqref{eq:WXX_WXL} and \eqref{eq:WLX_WLL}.

In this paper, we mainly make use of the operators $W_{XX}$ and $W_{XL}$.}

\edMK{\subsubsection{Identities}
Independently of the spacetime geometry, the BGOs satisfy identities related to the tangent vector $l^\mu$ \cite{Korzynski:2024jqt, Grasso:2018mei}:
\begin{eqnarray}
 W_{XX}(\lambda, \lambda_\mathcal{O})\,l_\mathcal{O} &=& l  \label{eq:identity1} \\
 W_{XL}(\lambda, \lambda_\mathcal{O})\,l_\mathcal{O} &=& (\lambda - \lambda_\mathcal{O})\,l \label{eq:identity2}\\
 W_{LX}(\lambda, \lambda_\mathcal{O})\,l_\mathcal{O} &=& 0 \label{eq:identity3} \\
 W_{LL}(\lambda, \lambda_\mathcal{O})\,l_\mathcal{O} &=& l .
 \label{eq:identity4}
\end{eqnarray}
In order to prove them, we first not that the vector function $\xi^\mu(\lambda) = A + B\,(\lambda - \lambda_\O)\,l^\mu$ with any constants $A, B$, i.e., the null tangent multiplied by a general linear function  of the affine parameter, is always a solution of \eqref{eq:GDE}. Substituting this to back to \eqref{eq:WXX_WXL} and \eqref{eq:WLX_WLL} we obtain the result.}

The perturbed geodesic defined by $\xi$ is null if we assume additionally that $\nabla_l \xi^\mu\,l_\mu = 0$.  Moreover, for null geodesics emitted and received at the same moment, an even stronger condition holds $\xi \cdot l = 0$, i.e., $\xi$ is orthogonal to $l$. The reader may check that if we assume the first condition at any point, it will be satisfied everywhere. If we assume them both to be satisfied at any point, they are both satisfied everywhere.

\subsubsection{Tetrads}
The GDE is easiest to solve if we express it in an appropriate tetrad. We consider tetrads for which the first vector $u^\mu$ defines the observer. The next two are the transverse $e_{\bm 1}^\mu, e_{\bm 2}^\mu$, and the last one is the tangent $l^\mu$. The orthogonality relations read
\begin{equation}
\begin{split}
u\cdot u &= -1, \\
u\cdot e_{\bm A} &= 0, \\
l\cdot e_{\bm A} &= 0, \\
e_{\bm A} \cdot e_{\bm B} &= \delta_{{\bm A}{\bm B}},
\label{eq:orthogonality_tetrad}
\end{split}
\end{equation}
and $l\cdot l =0 $ by definition. We do not assume anything about the product $l\cdot u$ except that it does not vanish. In this setup, any vector $\xi$ orthogonal to $l$ can be decomposed according to
$\xi = \xi^{\bm A}\,e_{\bm A} + \xi^{l}\,l$. We also note that for any other tetrad $(u',f_{\bm A'}, l)$ satisfying the orthogonality relations \eqref{eq:orthogonality_tetrad} , we have
$\xi^{\bm A'} = R^{\bm A'}{}_{\bm B'}\,\xi^{\bm B'}$ with $R$ denoting a $SO(2)$ rotation matrix. In particular, by a simple rotation of the two transverse vectors $f_{\bm A}$, we may ensure that the transverse components remain the same in both frames, despite the boost of the observer's vector from $u$ to $u'$. This fact gives the transverse components of vectors $\xi^\mu$ orthogonal to the direction of propagation $l$ an observer-independent and thus also tetrad-independent meaning. This observation does not apply to the component $\xi^{l}$ along the line of sight, which depends on the observer.

The tetrad constructed this way may be parallel propagated along the null ray, with all orthogonality conditions stated above preserved.

\subsection{Transverse 4-velocity difference}
Given the null vector $l$ defining the light propagation direction, the base 4-velocity $U$ and another 4-velocity $u$, it is possible to define  the transverse 4-velocity difference vector \edMK{between $u$ and $U$} in a covariant way, denoted by \edMK{$\mathcal{T}_{l,U}(u)$}, as
\beq
\mathcal{T}_{l,U}(u) = \frac{l\cdot U}{l\cdot u}\,u - U. \label{eq:Delta_u_def}
\eeq
\edMK{This vector represents the transverse velocity of of a source with velocity $u$ as measured in the frame of $U$.} Denoting the redshift between $U$ and $u$ by $z = \frac{l\cdot u}{l\cdot U} - 1$, we can also rewrite this expression as
\beq
\mathcal{T}_{l,U}(u) = \frac{1}{1+z}\,u - U.
\eeq
The vector $\mathcal{T}_{l,U}(u)$ is always orthogonal to $l$. Its transverse components $\mathcal{T}_{l,U}(u)^{\bm A}$ generalize the transverse components of the standard 3-velocity difference to relativistic motions. We may confirm this by linearizing the expression in the difference between $u$ and $U$ and noticing that it yields simply the transverse component of the velocity difference. Moreover, $\mathcal{T}_{l,U}(u)$ appears in the most general relativistic expressions for the position drift (proper motion) and the redshift drift, in the form of the difference between the observer 4-velocity and the source 4-velocity \cite{Korzynski:2024jqt, Grasso:2018mei, Korzynski:2017nas} \edMK{(it is denoted $\Delta u$ in these papers). The definition given here is covariant in the following sense}: it can be applied in any basis or frame, and on top of that, the transverse components $\mathcal{T}_{l,U}(u)^{\bm A}$ are invariant with respect to boosts of the frame in the sense defined in the previous section.

\edMK{The notion of the relativistic transverse velocity difference is not symmetric with respect to the swapping of $U$ and $u$. The reader may verify that in general ${\cal T}_{l,U}(u) \neq {\cal T}_{l,u}(U)$. }

\subsection{Geodesic deviation equation in the FLRW metric}
We assume a general FLRW metric as a background of the following form:
\beq
\d s^2 = -\d t^2 + a(t)^2\left(\d \chi^2 + S_k(\,\chi)^2\,\left(\d\theta^2 + \sin^2\theta\,\d \varphi^2\right)\right), \label{eq:FLRWmetric}
\eeq
with the curvature parameter $k = \pm 1,0$ and $S_k$ defined as
\beq
S_k(\chi) = 
\begin{cases}
\sin \chi & \textrm{if}\,k = 1, \\
\chi &\textrm{if}\, k = 0, \\
\sinh \chi &\textrm{if}\, k = -1. \label{eq:Sk_def}
\end{cases}
\eeq
We also define $C_k(\chi) = \frac{\d}{\d\chi}\,S_k(\chi)$:
\beq
C_k(\chi) = 
\begin{cases}
\cos \chi & \textrm{if}\,k = 1, \\ 
1 &\textrm{if}\, k = 0, \\
\cosh \chi &\textrm{if}\, k = -1. \label{eq:Ck_def}
\end{cases}
\eeq
We introduce the notation $U$ for the cosmic observer's 4-velocity,
$U^\mu = 
(1, 0, 0, 0)
$ in the coordinate tetrad.

Consider the observer $\O$ at the origin $\chi = 0$, receiving light at $t=t_\O$. Then a null geodesic passing through the observer, corresponding to a light ray observed at $\cal O$, parametrized by $\lambda$, takes the parametric form $\chi = \chi(\lambda)$, $t=t(\lambda)$, $\theta = \const$, $\varphi = \const$, with the functions $t(\lambda)$ and $\chi(\lambda)$ defined by appropriate integrals. Its tangent vector has the form of
\beq
\tilde l^\alpha = \frac{a_\O}{a(t)}\left(-U^\alpha + n^\alpha\right)\,\left(\tilde l_\O\cdot U_\O\right),
\eeq
where $n$ is a spatial vector defining the direction of propagation, $n\cdot n = 1$, $n\cdot U = 0$. The last factor ensures the correct normalization of $\tilde l$.  Note that in this paper, we consider geodesics parametrized backwards in time, from the observer to the source, consistent with the backward ray-tracing approach.

The geodesic deviation equation can be solved as well.  Let $\xi(\lambda)$ solve the GDE. Then, in a parallel-transported tetrad, we have
\bea
\label{eq:WXX_WXL2}
\xi(\lambda) &=& W_{XX}(\lambda,\lambda_\O)\,\delta x_\O + W_{XL}(\lambda, \lambda_\O)\,\Delta l_\O, \quad \quad
\\
\label{eq:WLX_WLL2} 
\nabla_{\tilde l}\xi(\lambda) &=& W_{LX}(\lambda,\lambda_\O)\,\delta x_\O + W_{LL}(\lambda, \lambda_\O)\,\Delta l_\O, \quad \quad
\eea
where $W_{XX}$, $W_{XL}$, $W_{LL}$, $W_{LX}$ are bilocal operators acting on vectors at $\O$ and producing vectors at point $\lambda$ along the null geodesic. In this paper, we consider solutions which preserve the null character of the geodesic. In the FOD approximation, this amounts to the condition $\Delta l \cdot \tilde l = 0$  imposed at $\O$. This condition is then satisfied at every point \cite{Grasso:2018mei, Korzynski:2024jqt}.

For the purpose of this paper, we need the two operators $W_{XL}$ and $W_{LL}$ in explicit forms for all pairs of points along a null geodesic in a FLRW spacetime, acting on vectors orthogonal to $\tilde l$. It turns out that they can be obtained using standard results of geometrical optics in an FLRW spacetime.

Let $\lambda_1$ and $\lambda_2$ denote two points on the null geodesic, \edMK{$\lambda_1$ being the light emission point and $\lambda_2$ the observation point in the future. We assume the geodesic to be parametrized backwards in time, from the observation point to the emission point, so $\lambda_1 > \lambda_2$}.  $W_{XL}$ from $\lambda_2$ to $\lambda_1$ is closely related to the Jacobi matrix of a light beam with a vertex at \edMK{the observation point} $\lambda=\lambda_2$  \cite{Grasso:2018mei,  PhysRevD.101.063506, Korzynski:2024jqt}, i. e. 
\bea
W_{XL}{}^{\bm A}{}_{\bm B}(\lambda_1, \lambda_2) &=& \delta^{\bm A}{}_{\bm B}\,D_{12}\,\left(\tilde l_2\cdot U_2\right)^{-1},\label{eq:WXL_AB}\\
W_{XL}(\lambda_1, \lambda_2)\,\tilde l_1 &=& (\lambda_1 - \lambda_2)\,\tilde l_1, \label{eq:WXL_tilde_l}
\eea
where $D_{12}$ is the angular diameter distance from 2 to 1 \footnote{The order of indices in the notation for the angular diameter distance is reversed with respect to the standard notation, i.e. $D_{12}$ denotes the angular diameter distance from the observer at 2 to the source at 1.}, while $\tilde l_1$ and $\tilde l_2$ denote the tangent vector $\tilde l$ at points $\lambda_1$ and $\lambda_2$ respectively. \edMK{The latter equation is simply the identity \eqref{eq:identity2}.}

From now on, we also use a shorthand notation for the BGOs from \edMK{the} point 2 to 1,
\beq
W_{XL}(\lambda_1, \lambda_2) \equiv W_{XL}^{1\leftarrow 2}.
\eeq
It is a  standard result that the angular diameter distance can be expressed in terms of the expansion history of the FLRW metric,
\beq\label{eq:angulardiameterdistance_from_chi}
D_{12} = a_1\,S_k(\chi_{12}),
\eeq
where the comoving distance from 1 to 2 is given by the integral over the scale factor $a$ \cite{PhysRevD.101.063506},
\beq\label{eq:chi_from_a} \chi(a_1, a_2) \equiv
\chi_{12} = \int_{a_1}^{a_2}\,\frac{\d a}{a^2\,H(a)},
\eeq
or, equivalently, over the redshift $z = \frac{a_2}{a}-1$,
\beq\label{eq:chi_from_z}
\chi_{12} = \int_{0}^{z_1}\,\frac{\d z}{H(z)}.
\eeq 

The $W_{LL}$ operator in turn may be obtained by differentiating $W_{XL}$ in a parallel-transported tetrad:
\beq
{W_{LL}}^{\bm i}{}_{\bm j}(\lambda_1, \lambda_2) = \frac{\partial}{\partial \lambda_1} {W_{XL}}^{\bm i}{}_{\bm j}(\lambda_1, \lambda_2).
\label{eq:WLLstart}
\eeq
For the full calculation of $W_{LL}$, we refer to Appendix \ref{app:WLL}. The final result we obtain is
\beq
W_{LL}{}^{\bm A}{}_{\bm B}(\lambda_1,\lambda_2) =\frac{a_2}{a_1}\,\Sigma(a_1,a_2)\,\delta^{\bm A}{}_{\bm B},
\label{eq:WLL_1}
\eeq
where we have introduced the short-hand notation,
\beq
\Sigma(a_1, a_2) = C_k\left(\chi\left(a_1, a_2\right)\right)- H_1\,a_1\,S_k(\chi\left(a_1, a_2\right)).
\label{eq:Sigmadef}
\eeq

\edMK{
In Appendix \ref{app:alternative}, we derive a slightly different expression for $\Sigma(a_1, a_2)$, in terms of the standard cosmological data: the spatial curvature parameter $\Omega_{k,0}$, the cosmologial redshift of  $a_1$, i. e. $1 + z_1 = \frac{a_0}{a_1}$, and the present value of the Hubble parameter $H_0$,
\beq
\Sigma(a_1, a_2) = \sqrt{1 - \Omega_{k,0}\,(H_0\,D_{12})^2\,(1+z_1)^2} 
- H_1\,D_{12}. \label{eq:SigmaFLRW}
\eeq
For models with negligible spatial curvature, this reduces to
\bea
\Sigma(a_1, a_2) =1 - H_1\,D_{12}.
\eea

}

\subsection{Lensing setup}
We assume that the source emits light at time $t_\S$, the light is deflected by the lens at $t_\LL$, while the observer receives it at $t_\O$, $t_\S < t_\LL < t_\O$. 
The scale factors corresponding to these moments are denoted $a_\S$, $a_\LL$, and $a_\O$, respectively.

\iffigures
\begin{figure} 
    \centering
    \includegraphics[width=0.7\linewidth]{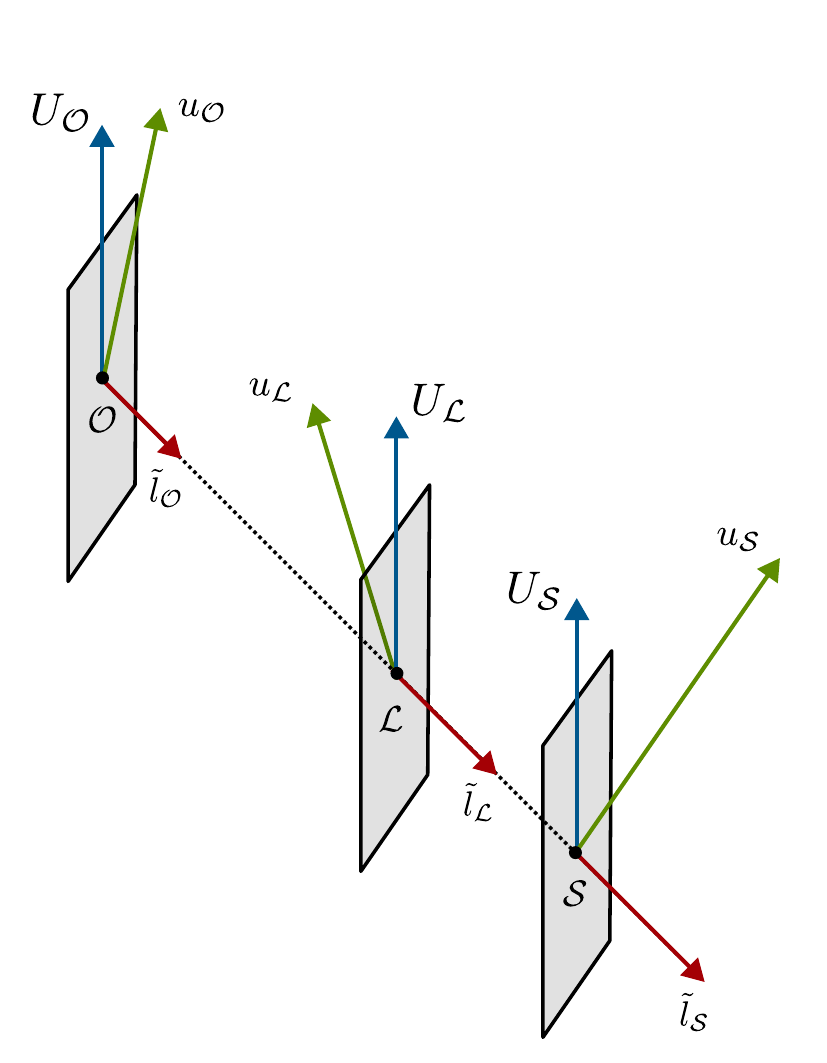}
    \caption{The lensing setup in a FLRW spacetime: the fiducial light ray, or the line of sight (dotted line), parametrized backwards in time and passing through the observer, lens, and source planes. The tangent vectors $\tilde l$ to the ray are denoted $\tilde l$, with the subscript denoting the point. We allow the 4-velocities of the observer $u_{\O}$, lens $u_{\LL}$, and source $u_{\S}$  to be different from the comoving cosmological observers $U_\O$, $U_\LL$, and $U_\S$, respectively. }
     \label{fig:lensing-setup-3d}
\end{figure}
\fi
\iffigures
\begin{figure} 
    \centering
    \includegraphics[width=0.8\linewidth]{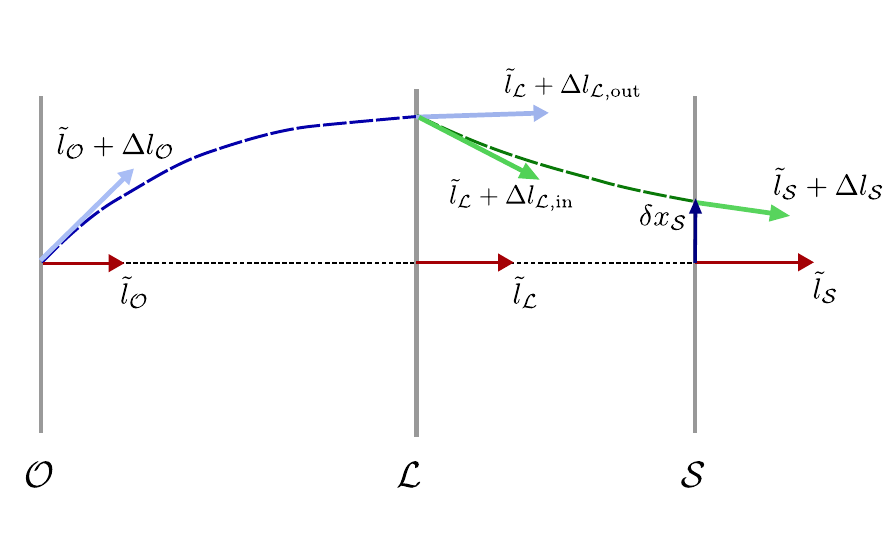}
    \caption{The lensed ray (dashed line) and the fiducial ray, or the line of sight (dotted line), as well as their tangent vectors at $\O$, $\LL$ and $\S$. The point source is located at $\delta x_\S$. The lensed ray undergoes a momentary deflection at the lens plane, which leads to a discontinuity of the tangent vector, represented by the pair $\tilde l_\LL + \Delta l_{\LL,\textrm{out}}$
    and $\tilde l_\LL + \Delta l_{\LL,\textrm{in}}$. The rays are parametrised backwards in time, from the observer to the source.} \label{fig:lensing-setup1-alt}
\end{figure}
\fi

Let $\tilde l$ denote the tangent to the unperturbed reference null geodesic passing through the observer, lens and source planes, see Figs. \ref{fig:lensing-setup-3d} and \ref{fig:lensing-setup1-alt}. The point $\O$ denotes the observation point, while $\LL$ and $\S$ are  fixed points on the lens and the source planes. For a static point source, we can identify $\S$ with the source's position, but for an extended or moving light source, we also need to consider luminous points on the lens plane located  off the reference geodesic. For that purpose, we introduce the notation $\delta x_\S^{\bm A}$ for the position on the source plane as measured from the reference point $\S$.

The 4-velocities of the source, the lens, and the observer, defining their momentary states of motion, are denoted by $u_\S$, $u_\LL$ and $u_O$. They are arbitrary future-oriented normalized timelike vectors. In particular, they do not need to coincide with the cosmic flow vectors at the points $\S$, $\LL$, and $\O$, denoted by $U_\S$, $U_\LL$, and $U_\O$.

Gravitational lensing is a result of the gravitational deflection, or transfer of 4-momentum to the ray at the lens plane, resulting in a discontinuity of the tangent vector to the ray.  We introduce the deflection 4-momentum transfer vector measuring this discontinuity,
\beq\label{eq:definition_defl}
\Delta l_{\textrm{defl}} = \Delta l_{\cal L,\textrm{out}} - \Delta l_{\cal L,\textrm{in}},
\eeq
where $\Delta l_{\cal L,\textrm{out}}$ and $\Delta l_{\cal L,\textrm{in}}$ are variations of 4-momentum on the lens plane after and before the deflection, respectively, see the  Fig. \ref{fig:lensing-setup1-alt}.

The angles defined in the standard lensing theory are related to the transverse components of $\Delta l_\O$, $\Delta l_\LL$ and $\Delta l_\S$. Since the frame of the observer is given by $u_\O$, the apparent position on the observer's sky is given by
\beq
\theta^{\bm A} = \frac{\Delta l^{\bm A}_\O}{\tilde l_\O\cdot u_\O}. \label{eq:theta_def}
\eeq
We also define the deviation angle as measured \emph{in the lens frame},
\beq
\widehat \alpha^{\bm A} = \frac{\Delta l^{\bm A}_\textrm{defl}}{\tilde l_\LL\cdot u_\LL}. \label{eq:widetildealpha_def}
\eeq
Finally, the variation of the viewing direction is defined \emph{in the source's frame},
\beq \label{eq:iota_def}
 \iota^{\bm A} = \frac{\Delta l^{\bm A}_\S}{\tilde l_\S\cdot u_\S}.
\eeq

\edMK{For small deviation angles the vectors $\Delta l_\O$, $\Delta l_{\rm defl}$ and $\Delta l_\S$ are all orthogonal to $\tilde l$. As we noted in Sec. \ref{sec:tetrads}, this means that the transverse components of $\Delta l_\O$, $\Delta l_{\rm defl}$ and $\Delta l_\S$ are frame independent (up to the rotation of the two transverse vectors), and thus, they are also well defined independently of the choice of the frame involved.

On the other hand, the angles $\theta, \widehat{\alpha}$ and $\iota$ do depend on the observer's, lens's and source's frames, but only via the denominators, which scale the angles up or down depending on the 4-velocities $u_\O$, $u_\LL$, and $u_\S$, respectively. This  dependence is due to the standard stellar aberration effect: the celestial sphere undergoes a conformal transformation under a boost of the reference frame. It follows that every small part of the celestial sphere, and also any angle defined on it, will appear conformally scaled up or down, depending on the frame, as reflected in Eqs. \eqref{eq:theta_def}-\eqref{eq:iota_def}.}

\subsection{Lensing equation} \label{sec:Lensing Equation}
We begin by rederiving the main result of the standard gravitational lensing theory, the lensing equation connecting the lensed and unlensed angle, with arbitrary velocities of the observer, lens and source. Let $\xi^\mu(\lambda)$  denote the solution of the geodesic deviation equation corresponding to the lensed ray. The ray undergoes a deflection at the lens plane, which in the thin lens approximation happens at a single point, leading to the discontinuity in the derivative $\nabla_l \xi^\mu$, but not in $\xi^\mu$.
\edMK{The GDE is linear, so we can obtain the position of the deflected ray by simply adding a correction term, denoted here $\xi_c$, to the undeflected result $\xi_u$,} i.e.
\beq
\xi(\lambda) = \xi_u(\lambda) + \xi_c(\lambda).
\eeq
The undeflected solution is given in terms of the BGOs by
\beq
\xi_u(\lambda) = W_{XL}(\lambda, \lambda_\O)\,\Delta l_{\O},
\eeq
since at $\O$ the ray crosses the origin $\delta x_{\O}=0$.
The deflection vanishes between the observer and lens plane $\lambda < \lambda_\LL$, while past the lens plane it does not vanish. It is easy to see that the correct expression is
\beq
\xi_c(\lambda) = \begin{cases}
  0 & \text{for }\lambda < \lambda_\LL, \\
  W_{XL}(\lambda,\lambda_\LL)\,\left(-\Delta l_{\text{defl}}\right) & \text{for }\lambda \ge \lambda_\LL.
\end{cases}
\eeq
With this choice, $\xi$ satisfies the geodesic deviation equation everywhere except the lens plane. It is unperturbed from the observer down to the lens, and it has the correct first derivative discontinuity at $\LL$.

We apply this formula to $\lambda = \lambda_\S$, obtaining
\beq
\delta x_\S  = W_{XL}^{\S\leftarrow\O}{}\,\Delta l_\O + W_{XL}^{\S\leftarrow\LL}\,\left(-\Delta l_\textrm{defl}
\right),\label{eq:DeltaxE}
\eeq
where we have introduced the short-hand notation for the BGOs from the observation point $\O$ to the source $\S$
\beq
W_{XL}^{\S\leftarrow\O} := W_{XL}(\lambda_\S, \lambda_\O),
\eeq
and the same for other pairs of points.
In a similar way, we can also derive the following relation for the derivative $\nabla_l \xi_\S$ at the emission point. In Sec. \ref{sec:tetrads}  we noted that $\nabla_l\xi\big|_\S \equiv \Delta l_\S$
\beq
\Delta l_\S  = W_{LL}^{\S\leftarrow\O}\,\Delta l_\O + W_{LL}^{\S\leftarrow\LL}\,\left(-\Delta l_\textrm{defl}\right)
\label{eq:DeltalE}.
\eeq

The lensing equation follows from the transverse components of \eqref{eq:DeltaxE}. Recall that  the transverse components of $W_{XL}$ are related to the angular diameter distances according to \eqref{eq:WXL_AB},
\bea
W_{XL}^{\S\leftarrow\O}{}^{\bm A}{}_{\bm B} &=& (\tilde l_{\O}\cdot U_\O)^{-1}\,D_{\S\O} \, \delta^{\bm A}{}_{\bm B}, \label{eq:deltaxe}
\\
W_{XL}^{\S\leftarrow\LL}{}^{\bm A}{}_{\bm B} &=& (\tilde l_{\LL}\cdot U_\LL)^{-1}\,D_{\S\LL} \, \delta^{\bm A}{}_{\bm B}, \label{eq:deltale}
\eea
where $D_{\S\O}$, $D_{\S\LL}$ is the angular diameter distance $\O$ to $\S$ and  $\O$ to $\LL$ , respectively, given by \eqref{eq:angulardiameterdistance_from_chi}. We also apply relations \eqref{eq:theta_def} and \eqref{eq:widetildealpha_def} to express $\Delta l_\O$ and $\Delta l_\text{defl}$ by the lensing angles $\theta$ and $\widehat \alpha$, obtaining
\beq
\delta x_\S^{\bm A} = \frac{\tilde l_\O\cdot u_\O}{\tilde l_\O \cdot U_\O}\,D_{\S\O}\,\theta^{\bm A} - \frac{\tilde l_\LL\cdot u_\LL}{\tilde l_\LL\cdot U_\LL}\,D_{\S\LL}\,\widehat \alpha^{\bm A}.
\eeq

This can be easily transformed into the proper lensing equation
\beq \label{eq:lensingequation}
\beta^{\bm A} = \theta^{\bm A} - \alpha^{\bm A},
\eeq
with the unlensed angle $\beta^{\bm A}$ defined in analogy to \eqref{eq:theta_def} as
\beq
 \delta x_\S^{\bm A} = {W_{XL}^{\S\leftarrow\O}}{}^{\bm A}{}_{\bm B} \,\left(\tilde l_\O\cdot u_\O\right)\,\beta^{\bm B},
\eeq
or, equivalently,
\beq
\delta x_\S^{\bm A} = \frac{\tilde l_\O\cdot u_\O}{\tilde l_\O \cdot U_\O}\,D_{\S\O}\,\beta^{\bm A},
\eeq
and the rescaled deflection angle $\alpha^{\bm A}$ defined by
\beq\label{eq:alphaA}
\alpha^{\bm A} = \frac{\tilde l_\LL\cdot u_\LL}{\tilde l_\LL\cdot U_\LL}\,\frac{\tilde l_\O\cdot U_\O}{\tilde l_\O\cdot u_\O}\,\frac{D_{\S\LL}}{D_{\S\O}}\,\widehat \alpha^{\bm A}. 
\eeq

\subsection{Frame-invariant approach to the frequency shift}

We now derive an important technical result: a simple and general formula for the variation of the redshift of an image due to the gravitational lensing, taking into account the effects of motions on the observer, lens and source, expressed in terms of the three angles
$\theta^{\bm A}$, $\alpha^{\bm A}$ and $\iota^{\bm A}$. The derivation is based on the assumption that in the vicinity of the lens, the metric is static to great accuracy, and thus we can assume that in the lens frame, the photon energy is conserved during the deflection. Note that we do not assume the Newtonian approximation to hold near the lens, unlike the standard derivations of the lensing theory. We only need the existence of a frame where the metric is static.

We have found it convenient in this context to make use of the logarithmic redshift $\zeta = \ln (1+z)$ rather than the standard one (a similar approach was  used in \cite{Ruggiero2006}). For a point source, it is given by the general formula 
\beq
\zeta = \ln(l_{\S}\cdot u_{\S}) - \ln(l_{\O}\cdot u_{\O})
\eeq
where $l_{\S} = \tilde l_{\S} + \Delta l_{\S}$, $\tilde l_{\S}$ is the unperturbed tangent vector and $\Delta l_{\S}$ is its variation due to the presence of the lens. 
This can be rearranged to
\begin{equation}
\begin{split}
\zeta = \ln\left(\frac{\tilde l_{\S}\cdot u_\S}{\tilde l_O\cdot u_\O}\right) & + 
\ln\left(1 + \frac{\Delta l_\S}{\tilde l_\S\cdot u_\S}\cdot u_\S\right) \\ & - \ln\left(1 + \frac{\Delta l_\O}{\tilde l_\O\cdot u_\O}\cdot u_\O\right).
\end{split}
\end{equation}
The first term is the unlensed logarithmic redshift, which we denote as $\zeta_0$. It is a constant for a given lens and source. 
We then have
\begin{equation}
\begin{split}
\zeta & = \zeta_0 + \Delta\zeta, \\
\Delta\zeta & =
\ln\left(1 + \frac{\Delta l_\S}{\tilde l_\S\cdot u_\S}\cdot u_\S\right) \\  & - \ln\left(1 + \frac{\Delta l_\O}{\tilde l_\O\cdot u_\O}\cdot u_\O\right).
\end{split}
\end{equation}
In this paper, we take the first order in deflection (FOD), borrowing the terminology from \cite{PhysRevD.69.063001}, linearizing this expression in $\Delta l$.
In the linear order, this amounts to
\beq \label{eq:deltazeta}
\Delta \zeta \equiv \frac{\Delta z}{1+z}=
 \frac{\Delta l_\S}{\tilde l_\S\cdot u_\S}\cdot u_\S - \frac{\Delta l_\O}{\tilde l_\O\cdot u_\O}\cdot u_\O.
\eeq
For $\Delta \zeta \ll 1 $, the redshift variation is related to the frequency shift $\Delta f$ of the wave via a simple formula $\frac{\Delta f}{f} = -\Delta \zeta$.

Both $\Delta l_\S$ and $\Delta l_\O$ are orthogonal to $\tilde l$ in the FOD approximation. Thus, both can be decomposed into the transverse components and the line-of-sight component,
\beq
\Delta l_\O = \Delta l_\O^{\bm A}\,e_{\bm A} + \edMK{\Delta l_\O^{\tilde l}}\,\tilde l,
\eeq
and a similar decomposition for $\Delta l_\S$
The transverse components are related to the angles $\theta$ and $\iota$ , respectively, see \eqref{eq:theta_def}-\eqref{eq:iota_def}. In the next step, we want to evaluate the expression \eqref{eq:deltazeta} in terms of the angles $\alpha$, $\theta$, and $\iota$. For this purpose, we need to relate the line-of-sight components $\Delta l_{\cal S}^{\tilde l}$ and $\Delta l_{\cal O}^{\tilde l}$ to the lensing angles with the help of the energy conservation at the lens plane.

We work in the parallel-transported adapted tetrad $\left(\widetilde {U_\LL}, e_{\bm A}, \tilde l\right)$, where $U_\LL\cdot e_{\bm A} = 0$, $e_{\bm A} \cdot e_{\bm B} = \delta_{\bm{AB}}$, $\tilde l\cdot e_{\bm A} = 0$. Here, $\widetilde{U_\LL}$ denotes parallel transport of $U_\LL$ to all points along the fiducial null geodesic.

The parallel-transported frame $\widetilde{U_\LL}$ at every point is related to the cosmological frame  $U$ by a boost along the line of sight. It follows then that the transverse components of vectors orthogonal to $\tilde l$ are identical in appropriately chosen tetrads related to $U$ and $\widetilde{U_\LL}$. We show this in detail below.

First, we consider the relationship between $\widetilde{U_\LL}$ and $U$ along the fiducial geodesic:
\bea
\widetilde{U_\LL} = a(\lambda)\,U + b(\lambda)\,\tilde l + c^{\bm A}(\lambda)\,e_{\bm A}.
\eea
We know that at the point $\LL$ we have simply $\widetilde{U_\LL} = U$, or $b = c^{\bm A} = 0$. It follows that $c^{\bm A} = 0$ everywhere. This can be seen by  solving directly the parallel transport equation for the FLRW metric \eqref{eq:FLRWmetric}, but it is also possible invoking a much simpler symmetry argument: the spacetime is perfectly symmetric with respect to the rotations around the fiducial null geodesic. These rotations are related to the rotations of the transverse vectors of the frame i. e. $e_{\bm A} \to R^{\bm B}{}_{\bm A}\,e_{\bm B}$, with $R^{\bm B}{}_{\bm A}$ being a 2D rotation matrix. Both the parallel transport equation and the initial data at $\LL$ obey thus rotational symmetry, so it follows that the transverse terms $c^{\bm A}$, which violate it, must vanish everywhere along the fiducial null geodesic.

It follows that at every point the tetrad $\left({U}, e_{\bm A}, \tilde l\right)$, obtained by replacing $\widetilde U_\LL$ by the local cosmic frame $U^\mu$ without affecting the rest, is also an adapted tetrad, i.e., the relation $e_{\bm A}\cdot \widetilde{U_{\LL}} = 0$ and other  orthogonality relations \eqref{eq:orthogonality_tetrad} still hold. We see that the boost in the direction of light propagation has no effect on the transverse vectors $e_{\bm A}$.

Take now a vector $\xi^\mu$ orthogonal to  $\tilde l$, $\xi^\mu\,\tilde l_\mu = 0$. In the parallel-transported tetrad $\left(\widetilde{U_\LL}, e_{\bm A}, \tilde l\right)$ its decomposition reads:
\beq
\xi^\mu = \xi^{\bm A}\,e_{\bm A}^\mu + \edMK{\xi^{\tilde l}}\,\tilde l^\mu, \label{eq:xidecomposition}
\eeq
i. e. it has no component along $\widetilde{U_\LL}$. However, note that the decomposition \eqref{eq:xidecomposition} works also for the other frame $\left({U}, e_{\bm A}, \tilde l\right)$, exactly because the two bases differ only by the vector that is not present in the decomposition.
The components $\xi^{\bm A}$ and $\xi^{\tilde l}$ are therefore identical in the parallel-transported tetrad $\big(\widetilde {U}_\LL, e_{\bm A}, \tilde l \, \big)$ and the local adapted tetrad 
$\big( {U}, e_{\bm A}, \tilde l \, \big)$. We may therefore think of these components as results of the decomposition in both the local adapted tetrad, and the parallel-transported tetrad.

The energy conservation takes the form of $\Delta l_{\LL,\textrm{out}}\cdot u_\LL = \Delta l_{\LL,\textrm{in}}\cdot u_\LL$, or, equivalently,
\beq\label{eq:Delta_l_u}
\Delta l_{\textrm{defl}}\cdot u_{\LL} = 0. 
\eeq
Obviously $\Delta l_\textrm{defl}\cdot \tilde l = 0$, so it has the decomposition 
\beq
\Delta l_{\textrm{defl}} = \widehat \alpha^{\bm A}\,(\tilde l \cdot u_{\cal L})\,e_{\bm A} + \edMK{\Delta l_{\textrm{defl}}^{\tilde l}\,\tilde l.}
\eeq
From this formula and  \eqref{eq:Delta_l_u}
we get 
\beq\label{eq:Delta_l_defl}
\edMK{\Delta l^{\tilde l}_\textrm{defl}} = -\widehat \alpha^{\bm A}\,u^{\cal L}_{\bm A}.
\eeq

We now move on to the $l$ components of the vectors $\Delta l_\O$ and $\Delta l_\S$. We take the component of \eqref{eq:DeltalE} along $\tilde l$. 
From \eqref{eq:identity3} we know that independently of the underlying geometry
 ${W_{LL}}^{\tilde l}{}_{\tilde l} = 1$, ${W_{LL}}^{\tilde l}{}_{\bm A} = 0$ in the parallel-transported tetrad.  Therefore, we obtain 
\beq \label{eq:lSl}
\Delta \edMK{l_{\cal S}^{\tilde l} = \Delta l_{\cal O}^{\tilde l} - \Delta l_{\textrm{defl}}^{\tilde l}}.
\eeq
Now, substituting \eqref{eq:Delta_l_defl} and \eqref{eq:lSl} to \eqref{eq:deltazeta}, we get 
\beq
\Delta \zeta = \frac{\Delta l_{\cal S}^{\bm A}}{\tilde l_{\cal S}\cdot u_{\cal S}}\,u^{\cal S}_{\bm A} + \widehat \alpha^{\bm A}\,u^{\cal L}_{\bm A} - \frac{\Delta l_{\cal O}^{\bm A}}{\tilde l_{\cal O}\cdot u_{\cal O}}\,u^{\cal O}_{\bm A}.
\eeq

Finally, we make use of \eqref{eq:theta_def} and \eqref{eq:iota_def} to substitute $\theta^{\bm A}$ and $\iota^{\bm A}$ for  the transverse components of $\Delta l_\O$ and $\Delta l_\S$.  In the end, we obtain a remarkably simple formula,
\highlight{
\beq \label{eq:Dz_from_angles}
\Delta \zeta = \iota^{\bm A}\,u^{\cal S}_{\bm A} + \widehat \alpha^{\bm A}\,u^{\cal L}_{\bm A} - \theta^{\bm A}\,u^{\cal O}_{\bm A}.
\eeq
}
We have found it useful  to express this formula in terms  of ``unphysical'' angles, measured in the cosmological comoving frames at each of the three points rather than the appropriate rest frames, i.e.,
\bea \label{eq:iotaU}
\edMK{\iota_{U_\S}^{\bm A}} &\equiv& \frac{\Delta l_{\cal S}^{\bm A}}{\tilde l_{\cal S}\cdot U_{\cal S}}=\frac{\tilde l_{\cal S}\cdot u_{\cal S}}{\tilde l_{\cal S}\cdot U_{\cal S}}\,\iota^{\bm A}, \\ \label{eq:thetaU}
\edMK{\theta_{U_\O}^{\bm A}} &\equiv& \frac{\Delta l_{\cal O}^{\bm A}}{\tilde l_{\cal O}\cdot U_{\cal O}}=\frac{\tilde l_{\cal O}\cdot u_{\cal O}}{\tilde l_{\cal O}\cdot U_{\cal O}}\,\theta^{\bm A}, \\
\label{eq:alphaU}
\edMK{\widehat \alpha^{\bm A}_{U_\LL}} &=&  \frac{\tilde l_{\cal L}\cdot u_{\cal L}}{\tilde l_{\cal L}\cdot U_{\cal L}}\,\widehat \alpha^{\bm A}. 
\eea
The angles are rescaled by Doppler prefactors with respect to those that have been measured at the observer, lens, or source frame. The formula for $\Delta \zeta$  takes the form of
\begin{equation}\label{eq:Dz_from_angles2}
\begin{split}
\Delta \zeta = \iota_{U_\S}^{\bm A}\,u^{\cal S}_{\bm A}\,\frac{\tilde l_{\cal S}\cdot U_{\cal S}}{\tilde l_{\cal S}\cdot u_{\cal S}} & + \widehat \alpha^{\bm A}_{U_\LL}\,u^{\cal L}_{\bm A}\,\frac{\tilde l_{\cal L}\cdot U_{\cal L}}{\tilde l_{\cal L}\cdot u_{\cal L}} \\ & - \theta^{\bm A}_{U_\O}\,u^{\cal O}_{\bm A}\,\frac{\tilde l_{\cal O}\cdot U_{\cal O}}{\tilde l_{\cal O}\cdot u_{\cal O}}.
\end{split}
\end{equation}
The advantage of this expression is that it can be easily recast into a manifestly frame-independent form, with the help of the transverse 4-velocity differences between the frames of the observer, lens and source and the appropriate cosmic frames. Namely, in the $\widetilde{U}_\O$ frame, we have  
\highlight{
\bea \nonumber
\Delta \zeta &=& \iota_{U_\S}^{\bm A} \, \edMK{\mathcal{T}_{\tilde l,U_{\S}} (u_{\S})_{\bm A}} + \widehat \alpha_{U_\LL}^{\bm A} \, \edMK{\mathcal{T}_{\tilde l,U_{\LL}} (u_{\LL})_{\bm A}} \\
&-& \theta_{U_\O}^{\bm A}\,\edMK{\mathcal{T}_{\tilde l,U_{\O}} (u_{\O})_{\bm A}},\label{eq:DZ_from_angles3}
\eea
} where each $\edMK{\mathcal{T}_{\tilde l,U_{*}} (u_{*})_{\bm A}}$ is the transverse velocity difference defined by \eqref{eq:Delta_u_def} between $u_*$ and $U_*$, with $*$ denoting $\O$, $\LL$ or $\S$. This is due to the fact that in this frame $U_\S$ and $U_\O$ have vanishing transverse components. For the sake of brevity, from now on we, use the short hand notation $\mathcal{T}(u_*) \equiv \mathcal{T}_{\tilde l,U_{*}} (u_{*})$ for the transverse velocity difference with the cosmological flow as the base 4-velocity.

Now, we note that the vectors $\mathcal{T}(u_*)$, just like $\Delta l_O$, $\Delta l_\S$, and $\Delta_\textrm{defl}$ are all orthogonal to $\tilde l$. It follows that each contraction of the transverse components in
\eqref{eq:DZ_from_angles3} is simply equal to the full spacetime scalar product of $\edMK{\mathcal{T}(u_{*})}$ with appropriate $\Delta l_*$,
\begin{equation}
\Delta \zeta = \frac{\Delta l_{\S} \cdot  \edMK{\mathcal{T}(u_{\S})}}{\tilde l_{\cal S}\cdot U_{\cal S}} + \frac{\Delta l_\textrm{defl}^{\LL}\cdot\edMK{\mathcal{T}(u_{\LL})}}{\tilde l_{\LL}\cdot U_{\LL}} - \frac{\Delta l_{\cal O}\cdot \edMK{\mathcal{T}(u_{\O})}}{\tilde l_{\cal O}\cdot U_{\cal O}}.
\end{equation}
This form of the formula for $\Delta \zeta$   is manifestly covariant and can be applied in any tetrad or any coordinate-based frame.

\section{Variation of the viewing direction} \label{sec:variation_of_the_viewing_direction}
The formulas derived in the previous section express $\Delta \zeta$ as a combination of terms involving all three angles $\theta^{\bm A}$, $\iota^{\bm A}$ and $\alpha^{\bm A}$. However, these angles are not independent: in the lensing theory, each ray is described by the apparent position on the observer's sky $\theta^{\bm A}$ and the deflection angle $\alpha^{\bm B}$. Therefore, in this section, we express the variation of the viewing direction angle $\iota^{\bm A}$ via the pair  $\theta^{\bm A}$ and $\alpha^{\bm A}$ in an FLRW universe model. Later, we trade $\alpha^{\bm A}$ for the unlensed angle $\beta^{\bm B}$, since the pair $\left(\theta^{\bm A}, \beta^{\bm B}\right)$ appears more useful from the point of view of the lensing theory.

Just like in the previous section, our method is based on the geodesic deviation equation and the BGO formalism.
We combine the transverse components of \eqref{eq:DeltalE} with the definitions of $\iota^{\bm A}$ \eqref{eq:iota_def},
$\theta^{\bm A}$ \eqref{eq:theta_def} and $\widehat \alpha^{\bm A}$ \eqref{eq:widetildealpha_def}, as well as the expression for $W_{LL}$ \eqref{eq:WLL_1}, and get
\begin{equation}
\begin{split}
\iota^{\bm A} & = \frac{\tilde l_\O\cdot u_\O}{\tilde  l_\S\cdot u_\S}\,\frac{a_\O}{a_\S}\,\Sigma(a_\S, a_\O)\,\theta^{\bm A} \\ & - \frac{\tilde l_\LL\cdot u_\LL}{\tilde l_\S\cdot u_\S}\,\frac{a_\LL}{a_\S}\,\Sigma(a_\S, a_\LL)\,\widehat\alpha^{\bm A}.\label{eq:iota1}
\end{split}
\end{equation} 
We then substitute $\widehat \alpha^{\bm A}$ by $\alpha^{\bm A}$ using  \eqref{eq:alphaA}, making use of $\frac{a_\LL}{a_\S} = \frac{\tilde l\cdot U_\S}{\tilde l\cdot U_\LL}$ to get
\begin{widetext}
\begin{equation}
 \frac{\tilde l_\LL\cdot u_\LL}{\tilde l_\S\cdot u_\S}\,\frac{a_\LL}{a_\S}\,\Sigma(a_\S, a_\LL)\,\widehat\alpha^{\bm A} = 
\frac{\tilde l_\S\cdot U_\S}{\tilde l_\S\cdot u_\S}\,\frac{\tilde l_\O\cdot u_\O}{\tilde l_\O\cdot U_\O}\,\Sigma(a_\S, a_\LL)\,\frac{D_{\S\O}}{D_{\S\LL}}\,\alpha^{\bm A}.
\end{equation}
\end{widetext}
After we substitute this into \eqref{eq:iota1}, it is easy to see that the Doppler prefactors in both terms on the right-hand side of \eqref{eq:iota1} are identical. We obtain
\highlight{
\beq \label{eq:iota2}
\iota^{\bm A} = \nu\left(\Sigma(a_\S, a_\O)\,\theta^{\bm A} - 
\Sigma(a_\S, a_\LL)\,\frac{D_{\S\O}}{D_{\S\LL}}\,\alpha^{\bm A}\right),
\eeq}
with the Doppler prefactor $\nu$ defined as
\beq \label{eq:nu_def}
\nu = \frac{\tilde l_\S\cdot U_\S}{\tilde l_\S\cdot u_\S}\,\frac{\tilde l_\O\cdot u_\O}{\tilde l_\O\cdot U_\O}.
\eeq
The prefactor $\nu$ can be absorbed if we use rescaled angles, defined in appropriate cosmological frames, instead of the standard ones, according to \eqref{eq:iotaU}-\eqref{eq:alphaU}. 
With these defintions, \eqref{eq:iota2}  can be rearranged to the following formula: 
\beq\label{eq:iota_from_theta_alpha}
\iota^{\bm A}_{U_\S} = \Sigma(a_\S, a_\O)\,\theta^{\bm A}_{U_\O} - 
\Sigma(a_\S, a_\LL)\,\frac{D_{\S\O}}{D_{\S\LL}}\,\alpha_{U_\O}^{\bm A},
\eeq
where we have introduced the rescaled angle $\alpha$ in the cosmological frame at $\O$,
\bea\label{eq:alpha_UO_def}
\alpha_{U_\O}^{\bm A} = \frac{\tilde l_\O\cdot u_\O}{\tilde l_\O\cdot U_\O}\,\alpha^{\bm A} . 
\eea

The formula \eqref{eq:iota_from_theta_alpha} becomes even more useful if we express it in terms of the angles $\theta_{U_\O}$ and the unlensed angle \edMK{ $\beta_{U_\O}$}, defining the position of the source. First, we define by analogy the unlensed angle in the cosmological frame  $U_\O$ 
\beq
\beta_{U_\O}^{\bm A} = \frac{\tilde l_\O\cdot u_\O}{\tilde l_\O\cdot U_\O}\,\beta^{\bm A}. \label{eq:betaUO}
\eeq 
We note that in this setting, the lensing formula \eqref{eq:lensingequation} is equivalent to 
\beq
\label{eq:lensingequation_UO}
\beta_{U_\O} = \theta_{U_\O} - \alpha_{U_\O}.
\eeq
After substituting this version of the lensing formula to \eqref{eq:iota_from_theta_alpha} , we get
\begin{equation}
\begin{split}
\iota^{\bm A}_{U_\S} & = \left( \Sigma(a_\S, a_\O) - \Sigma(a_\S, a_\LL)\,\frac{D_{\S\O}}{D_{\S\LL}} \right)\,\theta^{\bm A}_{U_\O} \\ & + 
\Sigma(a_\S, a_\LL)\,\frac{D_{\S\O}}{D_{\S\LL}}\,\beta_{U_\O}^{\bm A}.
\end{split}
\end{equation}
It turns out that the first term can be further simplified. In Appendix \ref{app:derivation} , we prove the following identity, valid in any FLRW spacetime:
\beq
D_{\S\LL}\,\Sigma(a_\S, a_\O) - D_{\S\O}\,\Sigma(a_\S,a_\LL) = -\frac{a_\S}{a_\LL}\,D_{\LL\O}. \label{eq:sigma_identity}
\eeq
It follows that
\highlight{
\begin{equation} \label{eq:iota_from_theta_betaU}
\iota^{\bm A}_{U_\S} = -\frac{a_\S}{a_\LL}\,\frac{D_{\LL\O}}{D_{\S\LL}}\,\theta^{\bm A}_{U_\O} + 
\Sigma(a_\S, a_\LL)\,\frac{D_{\S\O}}{D_{\S\LL}}\,\beta_{U_\O}^{\bm A}.
\end{equation}
}
This formula gives the viewing angle variations in terms of the lensed and unlensed angles. Passing to the physical angles $\iota^{\bm A}$, $\theta^{\bm A}$ and $\beta^{\bm A}$, defined in the proper frames of the source and the observer, is fairly straightforward using \eqref{eq:iotaU}-\eqref{eq:alphaU} and \eqref{eq:betaUO},
\highlight{
\begin{equation}\label{eq:iota_from_theta_beta}
\iota^{\bm A} = \nu\,\left(-\frac{a_\S}{a_\LL}\,\frac{D_{\LL\O}}{D_{\S\LL}}\,\theta^{\bm A} + 
\Sigma(a_\S, a_\LL)\,\frac{D_{\S\O}}{D_{\S\LL}}\,\beta^{\bm A}\right),
\end{equation}
}
with $\nu$ given by \eqref{eq:nu_def}.

\vspace{0.5mm}

\section{Frequency shift from lensing angles} \label{sec:spectroscopic_effect_of_lensing}
Having established all the necessary results in previous sections, we are now ready to derive the general formula for the redshift of the image in terms of the lensed angle $\theta$ and the unlensed angle $\beta$. 
The starting point is the Eq. \eqref{eq:DZ_from_angles3}, into which we substitute the relation \eqref{eq:iota_from_theta_betaU} for $\iota_{U_\S}$. For the $\widehat{\alpha}_{U_\LL}$ term, on the other hand, we need to use the rescaled version of the lensing equation \eqref{eq:lensingequation_UO}. 
We first note that combining \eqref{eq:alpha_UO_def}, \eqref{eq:alphaA} and \eqref{eq:alphaU}  yields the following relation: 
\beq
\alpha_{U_\O}^{\bm A} = \frac{D_{\S\LL}}{D_{\S\O}}\,\widehat\alpha^{\bm A}_{U_\LL}.
\eeq
Substituting now the lensing equation \eqref{eq:lensingequation_UO}  yields
\beq
\widehat\alpha^{\bm A}_{U_\LL} = \frac{D_{\S\O}}{D_{\S\LL}}\,\left(\theta_{U_\O}^{\bm A} - \beta_{U_\O}^{\bm A}\right).
\eeq
This formula, in turn, can be directly substituted into \eqref{eq:DZ_from_angles3}. The formula for $\Delta\zeta$ simplifies to
\highlight{
\begin{equation} \label{eq:DZ_from_angles4}
\begin{split}
\Delta \zeta & = \theta^{\bm A}_{U_\O}\,\left(-\edMK{\mathcal{T}(u_{\O})_{\bm A}} + \frac{D_{\S\O}}{D_{\S\LL}}\,\edMK{\mathcal{T}(u_\LL
)_{\bm A}} - \frac{a_\S}{a_\LL}\,\frac{D_{\LL\O}}{D_{\S\LL}}\,\edMK{\mathcal{T}(u_\S)_{\bm A}}
\right) \\
&+ \beta^{\bm A}_{U_\O}\,\frac{D_{\S\O}}{D_{\S\LL}} \left(\Sigma(a_\S,a_\LL)\,\edMK{\mathcal{T}(u_{\S})_{\bm A}} - \edMK{\mathcal{T}(u_{\LL})_{\bm A}}\right)
\end{split}
\end{equation}
}
Again, the angles $\theta_{U_\O}$ and $\beta_{U_\O}$ are related to the physical ones $\theta$ and $\beta$ by a simple rescaling
\eqref{eq:thetaU} and \eqref{eq:betaUO}:
\highlight{
\begin{widetext}
\begin{eqnarray} \label{eq:DZ_from_angles5}
\Delta \zeta &=&
\theta^{\bm A}\,\left(-\edMK{\mathcal{T}(u_{\O})_{\bm A}} + \frac{D_{\S\O}}{D_{\S\LL}}\,\edMK{\mathcal{T}(u_\LL
)_{\bm A}} - \frac{a_\S}{a_\LL}\,\frac{D_{\LL\O}}{D_{\S\LL}}\,\edMK{\mathcal{T}(u_\S)_{\bm A}}
\right)\,\frac{\tilde l_{\O}\cdot u_\O}{\tilde l_\O\cdot U_\O} 
\\
&+& \beta^{\bm A}\,\frac{D_{\S\O}}{D_{\S\LL}} \left(\Sigma(a_\S,a_\LL)\,\edMK{\mathcal{T}(u_{\S})_{\bm A}} - \edMK{\mathcal{T}(u_{\LL})_{\bm A}}\right)\,\frac{\tilde l_{\O}\cdot u_\O}{\tilde l_\O\cdot U_\O}. \nonumber
\end{eqnarray}
\end{widetext}
}
This formula generalizes Eq. (37) derived by Wucknitz and Sperhake in \cite{PhysRevD.69.063001} in the FOV approximation. We note that the first parenthesis in \eqref{eq:DZ_from_angles5} contains the same linear combination of transverse velocities as the Doppler velocity formula (37), up to the global sign reversal \footnote{The  sign reversal in the linear combination of velocities in the first term of \eqref{eq:DZ_from_angles5} in comparison to the corresponding expressions in \cite{PhysRevD.69.063001, Samsing:2024xlo} is related to differences in the definition of the actual measure of the spectroscopic effect. In \cite{PhysRevD.69.063001} the measure is the corresponding fictitious radial velocity, or  Doppler velocity, while in \cite{Samsing:2024xlo} the fundamental quantity is related to the variation of the frequency, not the redshift.}. More precisely, comparing (37) with the first term of \eqref{eq:DZ_from_angles5}, we see that the generalization of that formula to the fully relativistic case beyond the FOV is straightforward: the transverse 3-velocities need to be replaced with the relativistic transverse 4-velocity differences between the lens, observer or source and the comoving cosmological observer at the corresponding point. On top of that, we need a Doppler prefactor in front of the whole first term, depending mainly on the observer's velocity along the line of sight. It follows that if we linearize the first term in all velocities we recover the Wucknitz-Sperhake relation. 

Apart from the first term, formula \eqref{eq:DZ_from_angles5} contains the contribution from the unlensed angle $\beta^{\bm A}$. This term is absent in the analysis in \cite{PhysRevD.69.063001}, where the source is assumed to be positioned exactly along the line of sight, with $\beta^{\bm A} = 0$. It is responsible for the difference in redshift at different points of an extended luminous object.

We would like to point out the fundamental difference between the two terms in terms of the relation between the position of a point of the image and the redshift. The first term is obviously linear in the apparent position of the object $\theta^{\bm A}$, and this way it introduces an overall gradient of redshift across the observer's celestial sphere, affecting in the same way all images of a given source and every point within these images. The second term is linear in the angle $\beta^{\bm A}$, proportional to the actual position of the point on the source plane but related to the observed $\theta^{\bm A}$ in a complicated and usually nonlinear way.

\iffigures
\begin{figure}[htbp]
    \centering 
     \includegraphics[width=\linewidth]{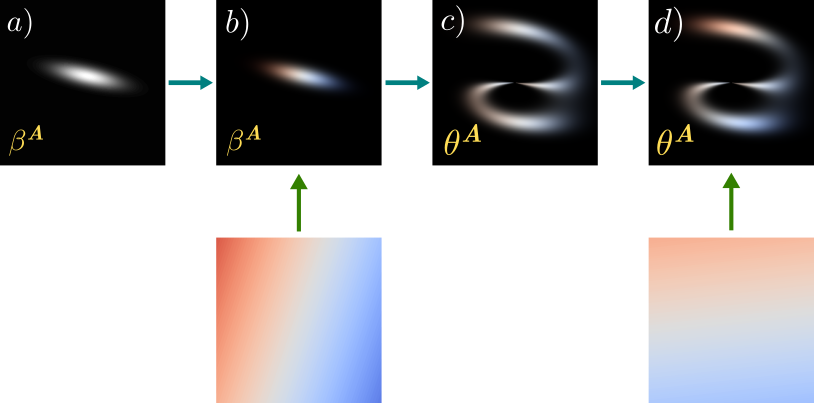}
    \caption{Decomposition of the frequency shift effect into three distinct transformations as described  by Eq. \eqref{eq:DZ_from_angles4}.  a)---b): Initial application of the frequency shift gradient, represented by the $\beta^{\bm A} $ term, across the unlensed source image. b)---c): The lensing map (modeled here as an isothermal ellipsoid) induces a nonlinear transformation, resulting in four partially overlapping images and a distorted frequency shift gradient. c)---d): Addition of a second frequency shift gradient on the lens plane, corresponding to the $\theta^{\bm A}$
  term. The hues of red and blue indicate redshifts and blueshifts, respectively.}
    \label{fig:3steps}
\end{figure}
\fi

We may therefore decompose the total spectroscopic effect of lensing into three distinct transformations applied successively, see  Fig. \ref{fig:3steps}. First, a gradient of frequency shift is applied to the \emph{unlensed} image on the source plane, with  magnitude and direction given by the second term of \eqref{eq:DZ_from_angles5}. The image is then mapped to the observer's plane using the lensing map \eqref{eq:lensingequation}. This step involves the distortion of the source image and possibly even splitting it into multiple images. It follows that the linear gradient of frequency shift from the first step is mapped to a nonlinear and potentially complicated dependence on the observed position of a point within each image. In the final, third step, another gradient of frequency shift is applied across the observed image, with the direction and magnitude given by the first term in \eqref{eq:DZ_from_angles5}. The two gradients are governed by  different combinations of the transverse velocities of the source, lens and observer. In particular, the combination standing in front of $\beta^{\bm A}$ does not depend at all on the observer's 4-velocity $u_{\O}$, while the combination in front of $\theta^{\bm A}$  depends on all three 4-velocities.

\iffigures
\begin{figure}[htbp]
    \centering 
     \includegraphics[width=\linewidth]{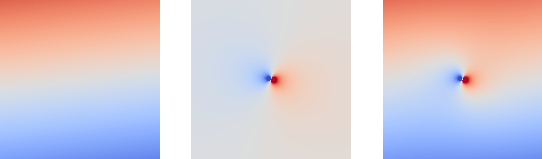}
    \caption{Frequency shift as a function of the lensed position, decomposed according to Eq. \eqref{eq:DZ_from_thetaalpha_UO} into the $\theta^{\bm A}_{U_\O}$ term in the form of a  gradient (left), the proper lensing term proportional to $\alpha_{U_\O}^{\bm A}$ (center), and the combined effect (right).
    The  gradient term is a long-range one, extending over the whole celestial sphere as a dipole, while the proper lensing term vanishes away from the lens. The lensing potential corresponds to a point mass, and the hues of red and blue indicate redshifts and blueshifts, respectively. }
    \label{fig:threeplots}
\end{figure}
\fi

If the spatial extent of the source is negligible, i. e. the source can be safely considered pointlike, and the temporal variation of its position with respect to the lens is negligible, then the first step is irrelevant. Only the gradient of redshift related to $\theta^{\bm A}$  contributes to the observed frequency of the images. The whole effect is parametrized by the two components in front of $\theta^{\bm A}$, which allows for the determination of the appropriate combination of transverse velocities using the redshifts of at least three images, a method known as the Doppler triangulation
\cite{Itoh.PhysRevD.80.044009, Samsing:2024xlo, PhysRevD.109.024064, Zwick_2025}. 

In  strongly lensing systems, the most prominent images are usually highly magnified, and therefore their angular size is significantly larger than the size of the unlensed image. Therefore, we may expect the contribution from the $\theta^{\bm A}$ term to be dominant over the $\beta^{\bm A}$ term in most cases. 

We may obtain even more  insights regarding the frequency effect and its dependence on relative motions of the source, lens and observer if we express formula \eqref{eq:DZ_from_angles4} via $\theta_{U_\O}^{\bm A}$ and the rescaled deflection angle $\alpha_{U_\O}^{\bm A}$ defined by \eqref{eq:alpha_UO_def}. Namely, using \eqref{eq:lensingequation_UO} and \eqref{eq:sigma_identity} , we can obtain:
\highlight{
\begin{equation}
\label{eq:DZ_from_thetaalpha_UO}
\begin{split}
\Delta\zeta & = \theta^{\bm A}_{U_\O}\,\left(-\edMK{\mathcal{T}(u_\O)_{\bm A}}
 + \Sigma\left(a_\S,a_\O\right)\,\edMK{\mathcal{T}(u_\S)_{\bm A}}\right) \\ & -\alpha^{\bm A}_{U_\O}\,\frac{D_{\S\O}}{D_{\S\LL}}\,\left(-\edMK{\mathcal{T}(u_\LL)_{\bm A}} + \Sigma\left(a_\S,a_\LL\right)\,\edMK{\mathcal{T}(u_\S)_{\bm A}}\right).
\end{split}
\end{equation}
}
As before, the angles $\theta^{\bm A}_{U_\O}$ and $\alpha^{\bm A}_{U_\O}$ can be substituted by their unrescaled counterparts, defined in the observer and the lens frame appropriately.

We note that this form of the equation appears much more symmetric than the previous ones. Moreover,  the total frequency shift has been decomposed into the effect of the observer-source relative motion, represented by the first term, and the proper frequency effect of lensing, proportional to the relative lens-source motion, given by the second term. The quantity representing the relative motion in both terms is not the difference of the transverse velocities of the two objects, but rather a  linear combination  involving the coefficient $\Sigma$. It coincides with the difference of velocities only if the lensing setup is located in a flat spacetime,  in which $\Sigma \equiv 1$ everywhere.

Just as before, the $\theta^{\bm A}_{U_\O}$ term defines a perfect dipole on the sky, and  is present even if there is no lens. It represents the variation of the Doppler frequency shift due to the change of line of sight as we compare two displaced but comoving sources. It is related to the secular decrease of the frequency of signals from pulsars moving transversely (Shklovskii effect) \cite{1970SvA....13..562S}. The  proper lensing  term, on the other hand, is strongly nonlinear as a function of $\theta^{\bm A}_{U_\O}$. However, since $\alpha^{\bm A}_{U_\O}$ is expected not to grow far away from the lens, its influence has only a limited range on the celestial sphere. Far away from the lens, it is  dominated by the dipole term, or even vanishes entirely, see Fig. \ref{fig:threeplots}.

Equation \eqref{eq:DZ_from_thetaalpha_UO} can be seen as a generalization of Eq. (6) from \cite{PhysRevD.101.024015}, in which we allow for relative motion between source and observers, arbitrary velocities, arbitrary lensing potential and distances for which the cosmological cannot be neglected.

\subsection{Estimate of the redshift difference between two images}

We now  estimate  the frequency effect of strong lensing in the case of a point source undergoing multiple imaging. In contrast to \cite{PhysRevD.101.024015}, we focus here on  line doubling and line broadening effect rather than the shift of the spectral barycenter of a single line. For electromagnetic radiation, the line broadening is the effect of unresolved multiple images with slightly offset frequencies.

Its  counterpart in the gravitational wave astronomy is the effect of frequency variation for two or more images. For  persistent sources, this manifests as  signal interference producing time-dependent modulation of the combined signal at the detector. For transient signals, such as binary inspirals, the effect may  be observed as a subtle time domain rescaling (stretching or squeezing) of the successive signals arriving at the detector  \cite{PhysRevD.109.024064, Itoh.PhysRevD.80.044009, Samsing:2024xlo}. 

Assume a point source at $\beta^{\bm A}$ with multiple images at $\theta_i^{\bm A}$. We would like to compare pairwise their redshift variations $\Delta \zeta_i$. Since they all share the same source position $\beta^{\bm A}$ , the second term from \eqref{eq:DZ_from_angles4} drops out, and we obtain for any pair of images $i$ and $j$
\begin{widetext}
\beq
\begin{split}
\Delta\zeta_{ij} & \equiv \Delta \zeta_i - \Delta \zeta_j, \\
&= \left(\theta_i^{\bm A}{}_{U_\O}-\theta_j^{\bm A}{}_{U_\O}\right)\,\left(-\edMK{\mathcal{T}(u_{\O})_{\bm A}} + \frac{D_{\S\O}}{D_{\S\LL}}\,\edMK{\mathcal{T}(u_{\LL})_{\bm A}} - \frac{a_\S}{a_\LL}\,\frac{D_{\LL\O}}{D_{\S\LL}}\,\edMK{\mathcal{T}(u_\S)_{\bm A}}\right).
\label{eq:Deltazetapairwise}
\end{split}
\eeq
\end{widetext}
This quantity is a product of two 2-vectors. It is maximized or minimised if the directions of both vectors coincide. In a general situation, we can still estimate the magnitude of the difference using the \emph{characteristic frequency variation} $\Delta\zeta_{\textrm{char}}$, 
\highlight{
\beq \label{eq:Deltazetapairwiseestimate}
\Delta\zeta _{\textrm{char}} = \Delta \theta_{ij}\,\left(\frac{\tilde l_\O\cdot u_\O}{\tilde l_\O\cdot U_\O}\right)\,v_{\textrm{eff}},
\eeq}
where $\Delta \theta_{ij}$ is the angular separation between the images $\left\|\theta_i - \theta_j\right\|$, and $v_\textrm{eff}$ is defined as the length of the transverse components of the effective transverse velocity:
\highlight{
\beq\label{eq:vDoppler}
v_{\textrm{eff}}^{\bm A} = -\edMK{\mathcal{T}(u_{\O})_{\bm A}} + \frac{D_{\S\O}}{D_{\S\LL}}\,\edMK{\mathcal{T}(u_{\LL})_{\bm A}} - \frac{a_\S}{a_\LL}\,\frac{D_{\LL\O}}{D_{\S\LL}}\,\edMK{\mathcal{T}(u_\S)_{\bm A}}
\eeq
}
In \cite{Samsing:2024xlo}, the sign reversal of this quantity is called the effective Doppler velocity; in \cite{PhysRevD.109.024064} the rescaled version of this quantity is called the effective lens velocity; in \cite{Itoh.PhysRevD.80.044009} , it has no name, while in \cite{PhysRevD.69.063001} the term Doppler velocity is used for the 
whole product \eqref{eq:Deltazetapairwiseestimate}.
Expression \eqref{eq:Deltazetapairwiseestimate} measures the characteristic scale of the variation of the redshift in a completely general lensing situation (compare Eq. (37) from \cite{PhysRevD.69.063001}) but without any restrictions regarding the 4-velocities of all three bodies involved.

\subsection{Point mass as a high velocity lens} \label{sec:pointmass_highvelocity}

As an example, we now consider a point mass $M$ with an arbitrary, potentially very high velocity as the lens. 
This case has already been discussed in \cite{PhysRevD.69.063001} in the context of the deflection angle: the authors noted that for lenses moving with relativistic speed toward the observer, the deflection is suppressed, and vanishes if the radial component of $v_\LL$ approaches $-1$. Here we focus on the spectroscopic effect and allow also for transverse motions with arbitrary velocity.

We neglect the velocities of the source and the observer and derive the estimate of the redshift difference between the two main images. We then discuss the frequency effect of lensing in the relativistic range ($v_{\LL}$ of the order of 1, but smaller) and ultrarelativistic limit ($v_{\LL} \to 1$). It turns out that the magnitude of the effect exhibits somewhat surprising dependence on the direction of motion and velocity.

The deflection angle in this case is given by the classic light deflection formula,
\beq
\widehat{\alpha}^{\bm A} = \frac{4GM}{b}\,n^{\bm A},
\eeq
where $n^{\bm A}$ denotes the axis connecting the center with the source's position, i.e., $n^{\bm A} = \frac{\beta^{\bm A}}{\beta}$, with $\beta = \sqrt{\beta^{\bm A}\,\beta^{\bm B}\,\delta_{\bm AB}}$ and
$\theta = \sqrt{\theta^{\bm A}\,\theta^{\bm B}\,\delta_{\bm AB}}$. The position of the ray at the lens plane  $\delta x^{\bm A}_\LL$ is given 
by
\beq
\delta x_\LL^{\bm A} = b\,n^{\bm A},
\eeq
where $b$ is the impact parameter. On the other hand, we have
\beq
\delta x_\LL^{\bm A} = {W_{XL}^{\LL\leftarrow\O}}^{\bm A}{}_{\bm B}\,\Delta l_\O^{\bm B}.
\eeq
We can connect $b$ with the angle $\theta$ by combining the two equations above with \eqref{eq:WXL_AB} and \eqref{eq:theta_def}, obtaining
\beq
b = \theta\,D_{\LL\O}\,\frac{\tilde l_\O\cdot u_\O}{\tilde l_\O\cdot U_\O},
\eeq
 where $\theta^{\bm A}=\theta\,n^{\bm A}$.
This in turn yields the following formula for the deflection angle:
\beq \label{eq:pointmass_alpha}
\widehat \alpha^{\bm A} = \frac{4GM}{\theta}\,\frac{\tilde l_\O\cdot U_\O}{\tilde l_\O\cdot u_\O}\,\frac{1}{D_{\LL\O}}\,n^{\bm A}.
\eeq
Substituting this into the Eq. \eqref{eq:alphaA} , we obtain the one-dimensional lensing equation for lens \eqref{eq:lensingequation} in the form of 
\beq \label{eq:pointmass_lensing}
\theta - \left(\frac{\tilde l_\LL\cdot u_\LL}{\tilde l_\LL\cdot U_\LL}\right)\,\left(\frac{\tilde l_\O\cdot U_\O}{\tilde l_\O\cdot u_\O}\right)^2\,\frac{D_{\S\LL}}{D_{\S\O}\,D_{\LL\O}}\,\frac{4GM}{\theta} = \beta.
\eeq
This means that the effective Einstein-Chwolson angle (radius), defined for arbitrary motions,  is given by
\beq \label{eq:thetaE}
\theta_E = \frac{\tilde l_\O\cdot U_\O}{\tilde l_\O\cdot u_\O}\,\sqrt{\frac{4GM\,D_{\S\LL}}{D_{\S\O}\,D_{\LL\O}}\,\frac{\tilde l_\LL\cdot u_\LL}{\tilde l_\LL\cdot U_\LL}},
\eeq
and the lensing equation takes the familiar form of
\beq
\beta = \theta - \frac{{\theta_E}^2}{\theta}.
\label{eq:lensing_equation_point_lens}
\eeq

Now, in this type of lens, we have two images and their typical separation of images in this type of system is twice the Einstein radius, i.e., $2\theta_E$. The characteristic redshift variation \eqref{eq:Deltazetapairwiseestimate} reads 
\beq
\Delta \zeta_{\textrm{char}} = 2\theta_E\,\left(\frac{\tilde l_\O\cdot u_\O}{\tilde l_\O\cdot U_\O}\right)\,v_{\textrm{eff}}.
\eeq
 Plugging this into Eqs. \eqref{eq:Deltazetapairwise} and \eqref{eq:vDoppler}, we obtain
\beq
v_{\textrm{eff}} = \frac{D_{\S\O}}{D_{\S\LL}}\,\left\|\edMK{\mathcal{T}(u_\LL)}\right\|,
\eeq
and
\beq
\Delta\zeta_{\textrm{char}}=  2\theta_E\,\left(\frac{\tilde l_\O\cdot u_\O}{\tilde l_\O\cdot U_\O}\right)\,\frac{D_{\S\O}}{D_{\S\LL}}\,\left\|\edMK{\mathcal{T}(u_\LL)}\right\|,
\eeq
for the redshift difference between the two images 1 and 2.

Finally, we substitute \eqref{eq:thetaE} to obtain the redshift difference in the form of
\beq
\Delta\zeta_{\textrm{char}} = 4\sqrt{\frac{GM\,D_{\S\O}}{D_{\S\LL}\,D_{\LL\O}}\,\frac{\tilde l_\LL\cdot u_\LL}{\tilde l_\LL\cdot U_\LL}}\left\|\edMK{\mathcal{T}(u_\LL)}\right\|.
\eeq
This can be expressed as a product of a velocity-invariant prefactor and a velocity-dependent coefficient. 
Introducing the decomposition
$u_\LL^\mu = \gamma_\LL\left(U_\LL^\mu + v_\LL^{\bm 3}\,e_{\bm 3}^\mu + v_\LL^{\bm A}\,e_{\bm A}^\mu\right)$,
 $\gamma_\LL = \left(1 - v_\LL^2\right)^{-1/2}$, we get
\beq
\begin{split}
\Delta\zeta_{\textrm{char}} &= 4\sqrt{\frac{GM\,D_{\S\O}}{D_{\LL\O}\,D_{\S\LL}}\cdot\gamma_\LL^{-1}\,\left(1+v_\LL^{\bm 3}\right)^{-1}}\,\gamma_\LL\,v_\LL^\textrm{tr}
\\
&= 4\sqrt{\frac{GM\,D_{\S\O}}{D_{\LL\O}\,D_{\S\LL}}}\,\frac{\gamma_\LL^{1/2}\,v_\LL^\textrm{tr}}{\left(1+v_\LL^{\bm 3}\right)^{1/2}},
\end{split}
\eeq
with the transverse velocity $v_\LL^\textrm{tr} = \sqrt{v_\LL^{\bm A}\,v_\LL^{\bm B}\,\delta_{\bm A\bm B}}$.

If $\alpha$ denotes the angle between the line of sight toward the observer and  the lens velocity vector ($v_{\LL}^\textrm{tr} = v_\LL\,\sin\alpha$, $0 \le \alpha \le \pi$), the characteristic redshift variation can be expressed as
\highlight{
\beq
\Delta\zeta_{\textrm{char}} =  4\sqrt{\frac{GM\,D_{\S\O}}{D_{\LL\O}\,D_{\S\LL}}}\,f(\alpha, v_\LL),
\eeq
with the function $f$ encapsulating the kinematical dependence of the result
\beq
f(\alpha, v_\LL) = \frac{\gamma_\LL^{1/2}\,v_\LL\,\sin\alpha}{\left(1-v_\LL\,\cos\alpha\right)^{1/2}}.
\eeq
}

The spectroscopic effect obviously vanishes for the motion directed along the line of sight, either toward the observer ($\alpha = 0$) or toward the source ($\alpha = \pi$).
Moreover, the kinematic factor $f$ for small velocities $v_\LL \ll 1$ can be approximated by $v_\LL\,\sin\alpha$, i.e., the transverse velocity $v^\textrm{tr}_\LL$. The dependence of the spectroscopic effect on the direction of motion follows a simple dipolar profile given by $\sin\alpha$. However, as we ramp up the velocity of the lens, both the Lorentz factor $\gamma_\LL$ and the denominator become significant, and the profile changes, see Fig. \ref{fig:fourplots}. The dependence becomes steeper in the vicinity of $\alpha = 0$, i.e. lens moving toward the observer. In fact, for relativistic velocities and $\alpha \neq 0$, i.e. $v_\LL \to 1$, the function $f$ can be factorized into
\beq
f(\alpha, v_\LL) \approx \gamma_\LL^{1/2}\,\,\sqrt{1 + \cos\alpha}.
\eeq
This approximate formula does not apply to a narrow range of angles around $\alpha = 0$, where the function has a steep but smooth profile connecting the maximal positive value and 0 for any finite $v_{\cal L}$,  see Figs. \ref{fig:fourplots} and \ref{fig:fourplots2}. Thus, for large velocities, the spectroscopic effect quickly vanishes at $\alpha = 0$ if the lens is moving toward the observer. On the other hand, for lenses receding from the observer, the dependence  on $\alpha$ has a more gradual slope. The vanishing of $f$ for $\alpha =0$ means that the frequency variation vanishes for every velocity, large or small, if the lens is moving directly toward the observer. Note that the same is true for the deflection angle, but only for large velocities \cite{PhysRevD.69.063001}.

\iffigures
\begin{figure}[htbp]
    \centering 
     \includegraphics[width=0.85\linewidth]{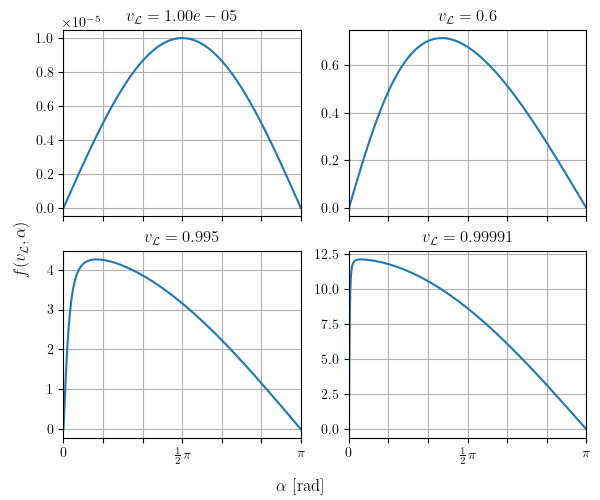}
    \caption{Dependence of the kinematical prefactor $f$ on the direction of the lens's motion, given by the angle $\alpha$, for the lens velocities $v_\LL$ of $10^{-5}$ (nonrelativistic, dipolar dependence on $\alpha$),  $0.6$ (mildly relativistic), $0.995$, and $0.99991$ (ultrarelativistic). Note that the scales on the $Y$ axes are different on each panel.}
    \label{fig:fourplots}
\end{figure}
\fi

Let us now turn to the problem of the magnitude of the spectroscopic effect. For a relativistic lens ($v_\LL$ comparable to $c = 1$, but smaller, say $0.7\,c$), the kinematical factor is at most of the order of $1$ and $\Delta \zeta_{\textrm{char}}$ is of the order of $4\sqrt{\frac{GM\,D_{\S\O}}{D_{\LL\O}\,D_{\S\LL}}}$. As we increase the velocity further, entering the ultrarelativistic regime, the $\gamma_\LL$ factor becomes dominant, and the characteristic redshift $\Delta \zeta_{\textrm{char}}$ grows, potentially out of bounds. However, we note that this growth is in practice very slow, at most of the order of $\gamma_\LL^{1/2}$. Because of the scaling with the square root of the Lorentz factor, even strong boosting of the lens increased the redshift by only a modest amount. 

\iffigures
\begin{figure}[htbp]
    \centering 
     \includegraphics[width= 0.85\linewidth]{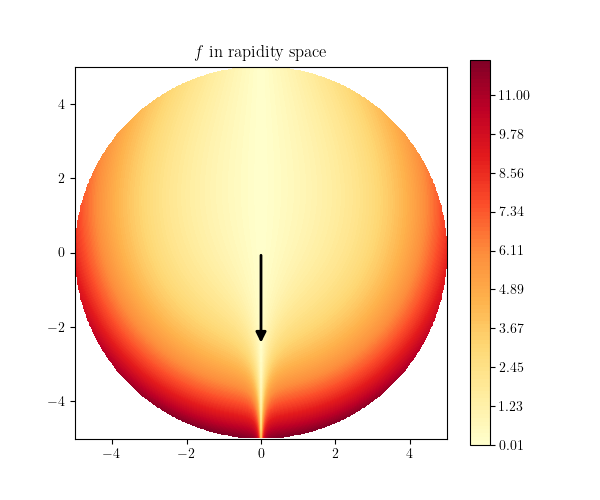}
    \caption{Kinematical prefactor $f$ in the rapidity space. The radial coordinate is the rapidity $w=\textrm{artanh}\,v_{\LL}$, while the polar coordinate is related to the angle $\alpha$ parametrizing the direction of the lens motion. The black arrow points toward the observer and $\alpha$ is defined as the angle from the axis marked by the arrow.}
    \label{fig:fourplots2}
\end{figure}
\fi

The distance dependence of the frequency variation is fairly simple: the spectroscopic effect for a moving lens and a stationary source and observer is large if either the lens is close to the source or to the observer. The minimal magnitude is attained for the lens located somewhere around the midpoint of the angular diameter distance. In the next section, we discuss both cases in more detail.

\subsection{Linearization in velocities}

For completeness, we present here the linearized versions of the formulas for the frequency variation and the viewing direction.
The expression we have derived  simplifies if we assume that the FOV approximation holds. In that case, we just need to replace the relative transverse 4-velocities by the transverse components of the 3-velocity and neglect the Doppler prefactors, which are all of the order of $1 + O(v)$. Thus, the Eq. \eqref{eq:iota_from_theta_beta} for the viewing direction takes the form of 
\highlight{
\beq\label{eq:iota_from_theta_beta_lin}
\iota^{\bm A} = -\frac{a_\S}{a_\LL}\,\frac{D_{\LL\O}}{D_{\S\LL}}\,\theta^{\bm A} + 
\Sigma(a_\S, a_\LL)\,\frac{D_{\S\O}}{D_{\S\LL}}\,\beta^{\bm A},
\eeq
}
while the equation for the redshift variation \eqref{eq:DZ_from_angles5} becomes
\highlight{
\beq
\begin{split} \label{eq:DZ_from_angles5_lin}
\Delta \zeta &=
\theta^{\bm A}\,\left(-v_{\O\,\bm A} + \frac{D_{\S\O}}{D_{\S\LL}}\,v_{\LL\,\bm A} - \frac{a_\S}{a_\LL}\,\frac{D_{\LL\O}}{D_{\S\LL}}\,v_{\S\,\bm A}\right)\,\\ 
&+ \beta^{\bm A}\,\frac{D_{\S\O}}{D_{\S\LL}} \left(\Sigma(a_\S,a_\LL)\,v_{\S\,\bm A} - v_{\LL\,\bm A}\right),
\end{split}
\eeq
}
with $v_*$ denoting the 3-velocity in the appropriate cosmological frame.

\section{Flat spacetime} \label{sec:flat_spacetime}

\subsection{Frequency shift in a flat spacetime}

A particular, but important case of an FLRW spacetime is the flat metric. It is appropriate as a physical model for the analysis of microlensing with the source and the lens in the immediate vicinity of the Galaxy. In this setting, $U^\mu$ denotes simply any constant timelike unit vector defining the frame in which we describe the lensing, so we can simplify the picture by moving to the observer's frame. Moreover, in flat spacetime, we have the additivity of distances, with $D_{\S\O} = D_{\S\LL} + D_{\LL\O}$.  This reduces the number of parameters describing the event by 2 to the two remaining distances (e.g. $D_{\S\LL}$ and $D_{\LL\O}$), the mass of the lens $M$ and the velocities of the lens and the source ($v_\LL$ and $v_\S$).

For a flat spacetime, the scale factor $a$ is always equal to 1, and the coefficients appearing in \eqref{eq:Sigmadef} simplify. Thus, the equation for the viewing direction variation \eqref{eq:iota_from_theta_betaU} takes the simplified form of
\highlight{
\beq
\iota_{U}^{\bm A} = -\frac{D_{\LL\O}}{D_{\S\LL}}\,\theta^{\bm A}_{U} + \frac{D_{\S\O}}{D_{\S\LL}}\,\beta^{\bm A}_{U},
\eeq
}
with $\iota_U$, $\beta_U$ and $\theta_U$ denoting the angles as measured in the $U$ frame. The formula for the spectroscopic effect of lensing \eqref{eq:Dz_from_angles} reads
\highlight{
\beq
\begin{split}
&\Delta\zeta = \\
& \theta^{\bm A}_{U}\,\left(-\edMK{\mathcal{T}_{l,U}(u_\O)}_{\bm A} + \frac{D_{\S\O}}{D_{\S\LL}}\,\edMK{\mathcal{T}_{l,U}(u_\LL)}_{\bm A} -\frac{D_{\LL\O}}{D_{\S\LL}}\,\edMK{\mathcal{T}_{l,U}(u_\S)}_{\bm A}\right) \\ 
&+ \beta^{\bm A}_{U}\,\frac{D_{\S\O}}{D_{\S\LL}} \left(\edMK{\mathcal{T}_{l,U}(u_\S)}_{\bm A} - \edMK{\mathcal{T}_{l,U}(u_\LL)}_{\bm A}\right).
\end{split} \label{eq:FlatDeltaZeta}
\eeq
}
In the Minkowski spacetime, we have the freedom to pick the reference frame $U^\mu$. An obvious choice is to align it with the frame of the observer $u_\O$ or the lens $u_\LL$, which further simplifies both expressions.

In the observer frame ($U^\mu = u_\O^\mu$) we get
\beq
\iota_{\O}^{\bm A} = -\frac{D_{\LL\O}}{D_{\S\LL}}\,\theta^{\bm A} + \frac{D_{\S\O}}{D_{\S\LL}}\,\beta^{\bm A}, \label{eq:flatspacetime_iota_O}
\eeq
i. e. we may use the observer's measured values of $\theta$ and $\beta$, and $\iota_\O \equiv \iota_{U_\S}$ is simply the viewing angle variation in the observer's frame. We also get
\beq
\begin{split}
\Delta \zeta &= \theta^{\bm A}\,\left( \frac{D_{\S\O}}{D_{\S\LL}}\,\edMK{\mathcal{T}_{l,u_{\O}}(u_\LL)}_{\bm A} -\frac{D_{\LL\O}}{D_{\S\LL}}\,\edMK{\mathcal{T}_{l,u_{\O}}(u_\S)}_{\bm A}\right)
\\ 
&+ \beta^{\bm A}\,\frac{D_{\S\O}}{D_{\S\LL}} \left(\edMK{\mathcal{T}_{l,u_{\O}}(u_\S)}_{\bm A} - \edMK{\mathcal{T}_{l,u_{\O}}(u_\LL)}_{\bm A}\right),
\end{split}
\eeq
with $\edMK{\mathcal{T}_{l,u_{\O}}(u_\LL)}$ and $\edMK{\mathcal{T}_{l,u_{\S}}(u_\LL)}$  being the transverse velocity difference between the observer and the lens and the observer and the source, respectively, see \eqref{eq:Delta_u_def}.

On the other hand, in the lens frame ($U^\mu = u_\LL^\mu$), we have
\beq
\begin{split}
\Delta \zeta &= -\theta^{\bm A}_{\LL}\,\left(\edMK{\mathcal{T}_{l,u_{\LL}}(u_\O)}_{\bm A} +\frac{D_{\LL\O}}{D_{\S\LL}}\,\edMK{\mathcal{T}_{l,u_{\LL}}(u_\S)}_{\bm A}\right)
\\ 
&+ \beta^{\bm A}_{\LL}\,\frac{D_{\S\O}}{D_{\S\LL}} \,\edMK{\mathcal{T}_{l,u_{\LL}}(u_\S)}_{\bm A} ,
\end{split}
\eeq
with the transverse velocity differences, this time calculated with respect to the lens frame.

Let us now discuss the redshift difference between two images, just like in \ref{sec:pointmass_highvelocity}. In the observer's frame we get
\beq
\Delta\zeta_{ij} = \Delta\theta_{ij}^{\bm A}\,\left(\frac{D_{\S\O}}{D_{\LL\S}} u_{\LL\,\bm A}
- \frac{D_{\LL\O}}{D_{\LL\S}}\,u_{\S\,\bm A}\right).
\eeq
We can now go back to the model of a point lens with a known Einstein radius from Sec. \ref{sec:pointmass_highvelocity}, assuming $\Delta\theta_{12}^{\bm A} = 2\theta_E\,n_{\bm A}$, where $n_{\bm A}$ is the unit vector indicating the direction of the line connecting the two images. Applying the formula for the Einstein-Chwolson angle  \eqref{eq:thetaE} and assuming the FOV approximation, we obtain (where we reintroduce the speed of light $c$),
\beq
\begin{split}
c\,\Delta\zeta_{12} & = 4\sqrt{\frac{GM}{c^2}\frac{D_{\S\LL}}{{D_{\S\O}\,D_{\LL\O}}}}\,\times \\ & \times \left(\frac{D_{\S\O}}{D_{\S\LL}} v_{\LL\,\bm A}\,n^{\bm A}
- \frac{D_{\LL\O}}{D_{\S\LL}}\,v_{\S\,\bm A}\,n^{\bm A}\right),
\end{split}
\eeq
and the corresponding characteristic redshift variation as
\highlight{
\begin{widetext}
\begin{equation}
\begin{split}
    c\, \Delta \zeta_{\text{char}} = \, & 2 \, v_{\text{eff}} \, \sqrt{\f{4GM}{c^2}\f{D_{\S\LL}}{D_{\S\O}D_{\LL\O}}} \\
     = \, & 830  \f{\text{cm}}{\text{s}} \,  \left( \f{(D_{\S\LL})/(D_{\S\O}D_{\LL\O})}{1 \text{ kpc}} \right) ^{\f{1}{2}} \left( \f{M}{M_\odot} \right)^{\f{1}{2}} \f{v_{\text{eff}}}{c} \\
     = \, & 0.28 \f{\text{cm}}{\text{s}} \, \left( \f{(D_{\S\LL})/(D_{\S\O}D_{\LL\O})}{1 \text{ kpc}} \right) ^{\f{1}{2}} \left( \f{M}{M_\odot} \right)^{\f{1}{2}} \f{v_{\text{eff}}}{100 \text{ km/s}},
     \label{eq:delta_z_char_dimensionful}
\end{split}
\end{equation}
\end{widetext}
}
where in the observer's frame $v_{\text{eff}}^{\bm{A}} = \left(\f{D_{\S\O}}{D_{\S\LL}}u_\LL^{\bm{A}} - \f{D_{\LL\O}}{D_{\S\LL}}u_\S^{\bm{A}}\right)$. The frequency variation is very small for velocities, masses, and distances typical of microlensing events in the Galactic neighborhood. Moreover, we can discuss three regimes when the formula \eqref{eq:delta_z_char_dimensionful} simplifies: lens very close the the source ($D_{\S\LL} \ll D_{\S\O}$, $D_{\S\O} \simeq D_{\LL\O}$), lens very close to the observer ($D_{\LL\O} \ll D_{\S\O}$, $D_{\S\LL} \simeq D_{\S\O}$) and the lens midway between the source and the observer ($D_{\S\LL} \simeq D_{\LL\O}$, $D_{\S\O} \simeq 2 D_{\LL\O} \simeq 2 D_{\S\LL}$). 

In the first case ($D_{\S\LL} \ll D_{\S\O}$), we get
\begin{equation}
\begin{split}
c\,\Delta \zeta_{\textrm{char}} & = 4\sqrt{\frac{GM}{c^2 D_{\S\LL}}}\,\left\|v_{\LL}^\textrm{tr} - v_{\S}^\textrm{tr}\right\| \\ 
& = 830  \f{\text{cm}}{\text{s}} \,  \left( \f{D_{\S\LL}}{1 \text{ kpc}} \right) ^{-\f{1}{2}} \left( \f{M}{M_\odot} \right)^{\f{1}{2}} \f{\left\|v_{\LL}^\textrm{tr} - v_{\S}^\textrm{tr}\right\|}{c}.
\end{split}
\end{equation}
Thus, the relevant velocity is the relative velocity between the lens and the observer. In the second case ($D_{\LL\O} \ll D_{\S\O}$), the first term in the parentheses dominates over the second one, i.e.,
\begin{equation}
\begin{split}
c\,\Delta \zeta_{\textrm{char}} & = 4\sqrt{\frac{GM}{c^2 D_{\LL\O}}}\,v_\LL^\textrm{tr} \\
& = 830 \f{\text{cm}}{\text{s}} \,  \left( \f{D_{\LL\O}}{1 \text{ kpc}} \right) ^{-\f{1}{2}} \left( \f{M}{M_\odot} \right)^{\f{1}{2}} \f{v_{\LL}^\textrm{tr}}{c}.
\end{split}
\end{equation}

In the third case ($D_{\S\LL} \simeq D_{\LL\O}$), we get
\begin{equation}
\begin{split}
c\,\Delta \zeta_{\textrm{char}} & = 4\sqrt{\frac{GM}{c^2 D_{\S\O}}}\,\left\|2v_{\LL}^\textrm{tr} - v_{\S}^\textrm{tr}\right\| \\ 
& = 830  \f{\text{cm}}{\text{s}} \,  \left( \f{D_{\S\O}}{1 \text{ kpc}} \right) ^{-\f{1}{2}} \left( \f{M}{M_\odot} \right)^{\f{1}{2}} \f{\left\|2v_{\LL}^\textrm{tr} - v_{\S}^\textrm{tr}\right\|}{c}.
\end{split}
\end{equation}
We can see that in both cases of lens close either to the source, or to the observer, the distance prefactors cancel to leave the smallest distance in the final denominator: either $D_{\S\LL}$ or $D_{\LL\O}$. This results in the amplification of the effect for such physical events  as a microlensing by a black hole passing near the Solar System, or lensing by a companion in a binary system (for example, binary black hole system orbiting a massive black hole), situation also known as self-lensing \edMKK{\cite{Ubach:2025oxr, 10.1111/j.1365-2966.2010.17490.x, Santos:2025ass}}. 

Let us also point out here that the presence of the prefactor $\Sigma$ in the Eqs. \eqref{eq:DZ_from_angles5} and \eqref{eq:DZ_from_thetaalpha_UO} for the redshift variation means that the equations for a lensing system in an FLRW spacetime cannot be obtained just by substituting the angular diameters into the relation \eqref{eq:FlatDeltaZeta} derived for the flat spacetime. Therefore, these prefactors constitute a genuinely new effect of the expanding spacetime present in gravitation lensing. This is in contrast to the  standard lensing equations from Sec. \ref{sec:Lensing Equation}, which are formally identical in both the Minkowski and FLRW spacetime when expressed via the angular diameter distance.

\subsubsection{Viewing direction variation in a flat spacetime}
\begin{figure}[tbp]
    \centering
    \includegraphics[width=0.85\linewidth]{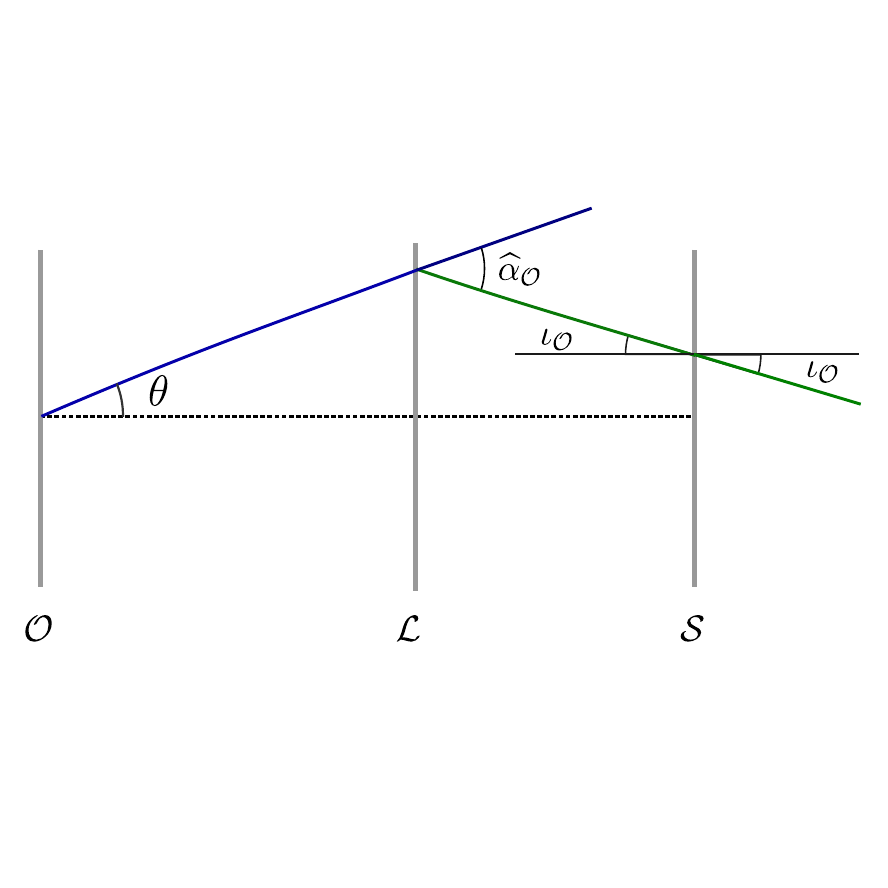}
    \caption{The angles $\theta$, $\widehat \alpha_{\cal O}$ and $\iota_{\cal O}$, defined in the observer's frame, in a flat spacetime. The plot shows the configuration as seen in the observer's frame while suppressing one spatial dimension and time.}
    \label{fig:lensing-flat}
\end{figure}

The viewing direction variation in a flat spacetime can be derived in a much simpler way using just planar geometry. The lensing setup, when viewed in the observer's frame ($U^\mu$ = $u_\O^\mu$), looks like in Fig. \ref{fig:lensing-flat}. It is easy to see from the geometry of angles involved that in this case we have a simple relation between the angles in the observer's frame,
\bea \label{eq:iotaO_theta_alpha}
\iota_\O^{\bm A} = \theta^{\bm A} - \widehat\alpha^{\bm A}_{\O}.
\eea
with $\widehat\alpha^{\bm A}_{\O} = \frac{\tilde l \cdot u_\LL}{\tilde l \cdot u_\O}\,\widehat\alpha^{\bm A}$ being the deflection angle measured in the observer's frame. 
From \eqref{eq:alphaA}, we then have
$\alpha^{\bm A} = \frac{D_{\S\LL}}{D_{\S\O}}\,\widehat \alpha_\O$, and combining this with the lensing equation \eqref{eq:lensingequation} and substituting the result as $\widehat\alpha^{\bm A}_{\O}$ to \eqref{eq:iotaO_theta_alpha}, we get immediately Eq. \eqref{eq:flatspacetime_iota_O}.

As a particular example, we may again consider the situation when there is no lens, or $\theta^{\bm A} = \beta^{\bm A}$. It follows that \eqref{eq:flatspacetime_iota_O} takes the form of 
\bea
\iota_{\O}^{\bm A} = \left(\frac{D_{\S\O} - D_{\LL\O}}{D_{\S\LL}}\right)\,\theta^{\bm A
},
\eea
and after applying the additivity of distances in a flat spacetime, we get $\iota_{\O}^{\bm A} = \theta^{\bm A}$; i. e., the viewing direction variation is exactly equal to the apparent position. This is because in this case, the rays being straight lines, the angle between the ray and the line of sight is strictly conserved. Thus, the lensed angle and the viewing direction variation, both measuring the angle between the ray and the line of sight on the observer and source plane respectively, are equal.

\subsection{Breaking of axial symmetry for axisymmetric lens}
\label{sec:breaking_symmetry}

\begin{figure}[htbp]
    \centering
    \includegraphics[width=0.7\linewidth]{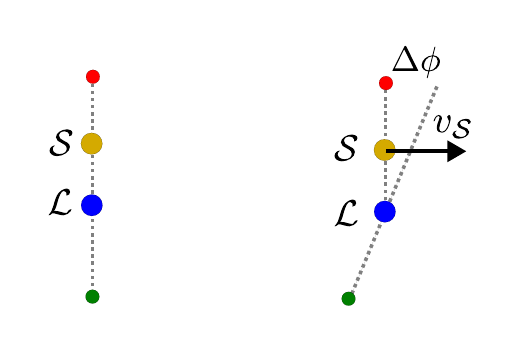}
    \caption{Position of different objects of a lensing system with an axisymmetric lens in the observer's sky. The yellow dot indicates the unlensed position of the source, the blue dot indicates the center of the lens, while red and green dots indicate the positions of the two images. Left: without any motion, the resulting images are collinear with the center of the lens. Right: when the source is moving behind the lens, the resulting images are no longer collinear with the center of the lens. We define the angle $\Delta \phi$ between the lines crossing the lens and each of the images.}
    \label{fig:asymmetric_images}
\end{figure}

We would like to point out one more effect happening for very fast lenses. Namely, due to significant time delay differences between the images, the perfect alignment between the two images and the lens is broken. In the stationary case with axisymmetric lensing potential (i.e., an axially or spherically symmetric lens, most notably microlensing by nonrotating stars and black holes), the images of a point source lie on a single line with both the source's unlensed position and the lens. This is no longer the case when the source is in motion behind the lens. As the emission times of different images are, in general, different, the corresponding source positions are different as well. We see that this results in the breaking of the perfect linear alignment as shown in Fig. \ref{fig:asymmetric_images}. It is important to stress, however, that this setup distinguishes between the motion of the source and the observer in the lens' frame: the motion of the observer incurs no additional time delay and thus does not contribute to breaking the axial symmetry of the images.

For concreteness, let us consider a point mass lens and a point source in a flat spacetime, with the lensing equation \eqref{eq:lensing_equation_point_lens}.   We aim to calculate the time delay between two images $\theta_{1,2}$ of the same source, with the lens being stationary with respect to the observer, while the source is moving with a nonrelativistic velocity $v_{\S,\textrm{tr}}^{\bm B}$. The positions of the images are the solutions to Eq. \eqref{eq:lensing_equation_point_lens} and read
\begin{equation}
    \theta_{1,2}^{\bm{A}} = \f{1}{2} \left(\beta \pm \sqrt{\beta^2 + 4 {\theta_E}^2} \right) \, n^{\bm{A}}.
    \label{eq:theta_point_mass}
\end{equation}
However, in writing so we have assumed that the position of the source of both images was the same, which was not actually the case. It becomes explicit once we calculate the time of arrival of both rays. The total time delay consists of the sum of the geometrical time delay according to the path traversed by the ray and the Shapiro delay due to the passing through a potential well and reads \cite{Blandford_Narayan_1986}
\begin{equation}
    \Delta T  = \f{D_{\LL\O} D_{\S\O}}{D_{\S\LL}} \left( \f{1}{2} \left( \theta^{\bm{A}} - \beta^{\bm{A}} \right)^2 - \Psi(\theta^{\bm{A}}) \right),
    \label{eq:time_delay}
\end{equation}
where $\Psi$ is the lensing potential of the lens.
\begin{figure}[tbp]
    \centering
    \includegraphics[width=0.7\linewidth]{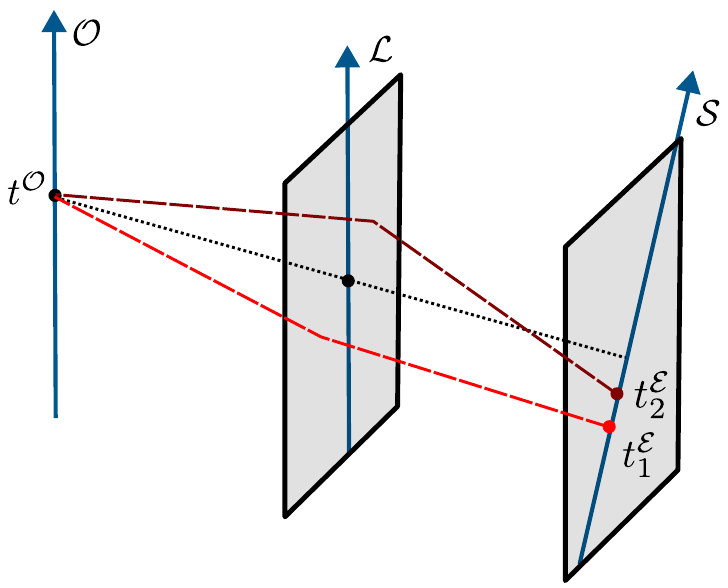}
    \caption{The lensing setup in a flat spacetime: the lens $\LL$ and observer $\O$ at rest, with the light source $\S$ in transverse motion. The time runs in the upward direction, and one spatial dimension has been suppressed. On the source's worldline, we have marked two different times of emission  $t^\E_{1,2}$ of light corresponding to the two images. Both of the light rays arrive at the observer at the same moment $t^\O$.}
    \label{fig:lensing-setup-3d-time_delay}
\end{figure}

The difference in times of emission of both light rays can be summarized in the implicit equation for the time of observation. Using the notation of Fig. \eqref{fig:lensing-setup-3d-time_delay}:
\begin{equation}
\begin{split}
    t^\O 
    & = t^\E_1 + T(\beta^{\bm{A}}(t^\E_1)) + \Delta T_1 (\beta^{\bm{A}}(t^\E_1)) \\
    & = t^\E_2 + T(\beta^{\bm{A}}(t^\E_2)) + \Delta T_2 (\beta^{\bm{A}}(t^\E_2)),
    \label{eq:time_delay_difference}
\end{split}
\end{equation}
where $t^\O$ is the time of observation, $t^\E_{1,2}$ are the times of emission of light coming through branches $1$ and $2$, respectively, $\beta^{\bm{A}}(t)$ is the position of the source at time $t$, $T(\beta^{\bm{A}})$ is the unlensed arrival time from  the source at $\beta^{\bm{A}}$, and $\Delta T_{1,2} (\beta^{\bm{A}})$ is the time delay of the light coming through branches $1$ and $2$, respectively, from the source at $\beta^{\bm{A}}$.

From this, we can calculate the difference in time of emissions, that is, the difference $\Delta t \equiv t^\E_2 - t^\E_1$,
\begin{widetext}
\begin{equation}
\begin{split}
    \Delta t \equiv t^\E_2 - t^\E_1 & = T (\beta^{\bm{A}}(t_1^\E)) + \Delta T_1 (\beta^{\bm{A}}(t_1^\E)) - T(\beta^{\bm{A}}(t_2^\E)) - \Delta T_2 (\beta^{\bm{A}}(t_2^\E))
    \\
    & = T(\beta^{\bm{A}}(t_1^\E)) + \Delta T_1 (\beta^{\bm{A}}(t_1^\E)) - T(\beta^{\bm{A}}(t_1^\E + \Delta t)) - \Delta T_2 (\beta^{\bm{A}}(t_1^\E + \Delta t)), 
\end{split}
\end{equation}
where the second-line expansion shows the presence of $\Delta t$-term in the rhs explicitly. We can solve this equation by expanding the $T$ and $\Delta T$ functions in Taylor series around $t^\E_1$,
\begin{equation}
\begin{split}
    T(\beta^{\bm{A}}(t^\E_1 + \Delta t)) & \simeq T \left( \beta^{\bm{A}}(t^\E_1) + \f{v_{\S,\text{tr}}^{\bm A}}{D_{\S\O}} \Delta t \right) \\ & \simeq T(\beta^{\bm{A}}(t^\E_1)) + \nabla_{\bm{B}}T(\beta^{\bm{A}}(t^\E_1)) \f{v_{\S,\text{tr}}^{\bm{B}}}{D_{\S\O}} \Delta t,
\end{split}
\end{equation}
and accordingly for the $\Delta T_2$ function. We have assumed here that the unlensed position $\beta^{\bm A}$ varies according to $\beta^{\bm A}(t) = \beta_0^{\bm A} + \frac{v_{\S,\textrm{tr}}^{\bm A}}{D_{\S\O}}\,t$, with $\beta_0^{\bm A}$ being a constant vector. The resulting difference in time of emissions reads
\begin{equation}
\begin{split}
    \Delta t & = \Delta T_1 (\beta^{\bm{A}}(t_1^\E)) 
    + \Big( \nabla_{\bm{B}} (\Delta T_2) (\beta^{\bm{A}}(t_1^\E)) + \nabla_{\bm{B}}(T)(\beta^{\bm{A}}(t_1^\E)) \Big) \cdot \f{v^{\bm B}_{\S,\text{tr}}}{D_{\LL\O}} \Delta t 
    - \Delta T_2 (\beta^{\bm{A}}(t_1^\E)).
\end{split}
\end{equation}
Solving this equation for $\Delta t$ yields
\begin{equation}
\begin{split}
    \Delta t & = \frac{\Delta T_1 (\beta^{\bm{A}}(t_1^\E)) - \Delta T_2 (\beta^{\bm{A}}(t_1^\E))}{1 - \left( \nabla_{\bm{B}} (\Delta T_2) (\beta^{\bm{A}}(t_1^\E)) - \nabla_{\bm{B}}(T)(\beta^{\bm{A}}(t_1^\E)) \right) \cdot \f{v^{\bm {B}}_{\S,\text{tr}}}{D_{\LL\O}}}
    \\ & \simeq \left( \Delta T_1 (\beta^{\bm{A}}(t_1^\E)) - \Delta T_2 (\beta^{\bm{A}}(t_1^\E)) \right) \left(1 + \left( \nabla_{\bm{B}} (\Delta T_2) (\beta^{\bm{A}}(t_1^\E)) + \nabla_{\bm{B}}(T)(\beta^{\bm{A}}(t_1^\E)) \right) \cdot \f{v^{\bm{B}}_{\S,\text{tr}}}{D_{\LL\O}} \right).
\end{split}
\label{eq:delay_moving}
\end{equation}
\end{widetext}
This complicated expression gives the difference in time of emissions between two branches, taking into account both the different time delays of different branches, as well as the change of the unlensed arrival time with the motion of the source. We can, however, notice that the main contribution to the expression comes from the difference between $\Delta T_1$ and $\Delta T_2$ functions, i.e., that light arrives via different branches of rays, as the second term in the parenthesis involves the transverse velocity $v_{\S,\textrm{tr}}^{\bm B}$ and is thus subdominant. We can thus approximate the difference in time of emissions by
\begin{equation}
    \Delta t \simeq \Delta T_1(\beta^{\bm{A}}(t_1^\E) - \Delta T_2(\beta^{\bm{A}}(t_1^\E)).
\label{eq:time_of_emission_difference}
\end{equation}

The angle of asymmetry is related to the difference in time of emissions by the effective angular velocity of the two images revolving around the lens on the observer's sky. It is straightforward that for a source moving with a constant velocity behind a lens with an impact parameter $b$ the angular velocity at the moment of closest approach to the lens is
\begin{equation}
    \omega = \f{v_{\S,\text{tr}}}{b \, D_{\S\O}}.
\end{equation}
Accordingly, we can calculate the angle of asymmetry in the image by
\begin{equation}
    \Delta \phi_{\text{max}} = \f{v_{\S\text{,tr}}}{b \, D_{\S\O}} \Delta t.
\end{equation}
We can estimate the order of the time delay by setting $\beta = b$ in Eqs. \eqref{eq:theta_point_mass} and \eqref{eq:time_delay}, calculating the time delay as if the source was positioned at the closest point to the lens.
For a full derivation, we refer to Appendix \ref{app:asymmetry_angle}.
It yields the maximal asymmetry angle as (where we again reintroduce the speed of light $c$ and assume that the impact parameter is small $b/\theta_E \ll 1$ and thus $\theta_1 \approx \theta_2$)
\begin{equation}
    \Delta \phi_{\text{max}} = \f{1}{2} \f{v}{c} \f{D_{\O\LL}}{D_{\LL\S}} \sqrt{b^2 + 4\theta_E^{\,\,2}}.
\end{equation}
We can see that this effect is strongly suppressed by the speed of light $c$, and for typical astrophysical values $v \approx 100 \text{ km/s}$, $b \approx \theta_E \approx \text{mas}$, the angle will be of the order of at most microarcseconds, far below the current observational resolution for single images.

\section{Summary and outlook} \label{sec:summary}

In this paper, we have presented a detailed derivation of the frequency effect of strong lensing in the most general situation, assuming a single lens in the thin lens approximation and any states of motion for the observer, lens and source. We have shown that the full spectroscopic effect can be expressed as a sum of two terms depending on different angles and discussed their dependence on the velocities of source.

In the electromagnetic domain, the frequency variation results in simultaneous splitting or broadening of spectral lines. In the gravitational wave domain, the result is the emergence of an interference pattern in the signal, both governed by the relative redshift difference $\Delta\zeta$.

Summarizing the findings of Secs. \ref{sec:spectroscopic_effect_of_lensing} and \ref{sec:flat_spacetime}, we note that the general frequency effect of lensing scales like
\beq
\Delta\zeta \approx 4 \sqrt{\frac{GM}{c^2\,D_{\textrm{eff}}}}\,\frac{v_{\textrm{eff}}}{c},
\eeq
with $M$ being the mass of the lens, $v_\textrm{eff}$ being the appropriately defined velocity scale, given by a combination of the velocities of the observer, lens and source, and $D_\textrm{eff}$ being the relevant distance scale, and we have explicitly reintroduced the speed of light $c$. In practice, $D_\textrm{eff}$ is the smaller of the two distances: the one between the lens and the source and between the lens and the observer. 

It follows that the frequency effect of lensing is small in most astrophysical situations, since it is expressible as a product of two small numbers: the transverse velocity expressed in units of the speed of light and the square root of the ratio of the lens's Schwarzschild radius to the distance scale. In fact, from \eqref{eq:DZ_from_angles5} it is obvious that the it can be large only in certain circumstances: if the distance between the images $\theta$ is large, i. e. the lens is fairly close to the observer, if it is very massive, or if the distance between the lens and the source $D_{\S\LL}$ is very small (small denominators). The latter situation is possible if the source is gravitationally bound to the lens in a binary system (self-lensing). As an example, consider a source orbiting a massive black hole on a fairly low orbit. The velocity can be fairly large in this case, i.e., close to $c$, and for a low orbit, the square root is likely to be large as well ($D_{\LL\S}$ of the order of up to a few Schwarzschild radii), and $\Delta\zeta$ can reach values of 1. If the source happens to pass periodically behind the BH on an orbit of small orbital inclination, we would have a chance to measure a pronounced line doubling effect due to the lensing on the central mass. Recently, this scenario has been discussed in the context of gravitational waves emitted by compact coalescing binaries in \cite{Ubach:2025oxr}.

\begin{acknowledgments}

The work of the first author was funded by the National Science Centre, Poland (NCN) under the Weave-UNISONO call (Grant No. 2024/06/Y/ST2/00190), for the joint Polish-Austrian project \emph{``Lensing of electromagnetic and gravitational waves''}. This research was funded in whole or in part by the Austrian Science Fund (FWF) \href{https://doi.org/10.55776/PIN9589124}{10.55776/PIN9589124}.

The work of the second author was funded by the National Science Centre, Poland, through the OPUS Project No. 2023/49/B/ST2/02782. For the purpose of Open Access, the authors have applied a CC-BY public copyright licence to any Author Accepted Manuscript version arising from this submission.

The authors thank Sudhagar Suyampraksam, Helena Ubach, Johan Samsing, Micha\l{} Bejger and Łukasz Wyrzykowski for their insightful discussions and comments.

\end{acknowledgments}

\appendix

\section{DERIVATION OF \texorpdfstring{$W_{LL}$}{WLL}}
\label{app:WLL}

In this appendix, we derive formula \eqref{eq:WLL_1}. We begin with formula \eqref{eq:WLLstart} for $W_{LL}$,
\bea
{W_{LL}}^{\bm i}{}_{\bm j}(\lambda_1, \lambda_2) = \frac{\partial}{\partial \lambda_1} {W_{XL}}^{\bm i}{}_{\bm j}(\lambda_1, \lambda_2),
\eea
along with the formulas \eqref{eq:WXL_tilde_l} for $W_{XL}$,
\bea
W_{XL}{}^{\bm A}{}_{\bm B}(\lambda_1, \lambda_2) &=& \delta^{\bm A}{}_{\bm B}\,D_{12}\,\left(\tilde l_2\cdot U_2\right)^{-1},
\label{eq:WXL_AB_app}
\\
W_{XL}(\lambda_1, \lambda_2)\,\tilde l_1 &=& (\lambda_1 - \lambda_2)\,\tilde l_1. \label{eq:WXL_tilde_l_app}
\eea
By differentiating \eqref{eq:WXL_tilde_l_app}, we have
\bea 
W_{LL}(\lambda_1, \lambda_2)\,\tilde l_2 = \tilde l_1.
\eea
For the transverse components, we need to take the derivative of \eqref{eq:WXL_AB_app}.
Let $D(t_1, t_2)$ denote the angular diameter distance $D_{12}$ as a function of cosmic time of emission $t_1$ and reception $t_2$. Differentiating the appropriate distances \eqref{eq:chi_from_a} and \eqref{eq:angulardiameterdistance_from_chi}, we get
\beq
\frac{\partial\,D\left(t(\lambda_1), t(\lambda_2)\right)}{\partial \lambda_1}= \frac{\partial}{\partial t_1} \,D\Big|_{t_1,t_2}\,\frac{dt}{d\lambda}\Big|_{\lambda=\lambda_1}.
\eeq
Obviously,
\beq
\frac{dt}{d\lambda}\Big|_{\lambda=\lambda_1} = \tilde l^t(\lambda_1).
\eeq
On the other hand, 
differentiating \eqref{eq:angulardiameterdistance_from_chi} with respect to $t_1$ yields
\beq
\label{eq:ddA}
\frac{\partial}{\partial t_1}D(t_1, t_2) = H_1\,a_1\,S_k(\chi_{12}) +
a_1 \,C_k\,\frac{\partial}{\partial t_1} \chi(t_1,t_2),
\eeq
with $\chi_{12} \equiv \chi(t_1, t_2)$ as a function of the emission and reception times. In \eqref{eq:chi_from_a} $\chi$ is defined via an integral with respect to $a$. Thus, we may reexpress $\chi_{12}$ by the scale factors $(a_1, a_2)$ rather than the cosmic times $(t_1, t_2)$.
We can then compute the derivative of $\chi$,
\beq
\begin{split}
\frac{\partial}{\partial t_1} \chi(t_1,t_2) & = 
\frac{da}{dt}\Big|_{t=t_1} \,\frac{\partial}{\partial a_1} \chi(a_1,a_2) \\
& = H_1\,a_1\,\frac{\partial}{\partial a_1} \chi(a_1,a_2).
\end{split}
\eeq
From \eqref{eq:chi_from_a}, we obtain 
\beq
\frac{\partial}{\partial a_1} \chi(a_1,a_2) = -\frac{1}{a_1^2\,H_1}.
\eeq
Substituting these relations back to \eqref{eq:ddA} yields
\beq
\frac{\partial}{\partial t_1}D(t_1, t_2) = H_1\,a_1\,S_k(\chi(t_1, t_2)) -
C_k( \chi(t_1,t_2)).
\eeq
From that, we get
\begin{widetext}
\beq
\begin{split}
W_{LL}{}^{\bm A}{}_{\bm B}(\lambda_1,\lambda_2) &=
\frac{1}{\tilde l_2\cdot U_2}\,\left(H_1\,a_1\,S_k(\chi_{12}) - C_k(\chi_{12})\right)\,l^t(\lambda_1) \,\delta^{\bm A}{}_{\bm B} \\
&= -\frac{\tilde l_1\cdot U_1}{l_2\cdot U_2}\left(H_1\,a_1\,S_k(\chi_{12}) - C_k(\chi_{12})\right)\,\delta^{\bm A}{}_{\bm B}.
\end{split}
\eeq
\end{widetext}
Recalling that $\frac{\tilde l_1\cdot U_1}{\tilde l_2\cdot U_2} = \frac{a_2}{a_1}$, we get the final result,
\beq
W_{LL}{}^{\bm A}{}_{\bm B}(\lambda_1,\lambda_2) = 
-\frac{a_2}{a_1}\left(H_1\,a_1\,S_k(\chi_{12}) - C_k(\chi_{12})\right)\,\delta^{\bm A}{}_{\bm B}.
\eeq

\edMK{
\section{DERIVATION OF THE ALTERNATIVE FORMULA FOR \texorpdfstring{$\Sigma(a_1, a_2)$}{Sig}  \texorpdfstring{\eqref{eq:SigmaFLRW}}{(\ref{eq:SigmaFLRW})} } \label{app:alternative}

From the definitions of $C_k(\chi)$ and $S_k(\chi)$ \eqref{eq:Sk_def} and\eqref{eq:Ck_def}, it follows that $C_k(\chi)^2 + k\,S_k(\chi)^2 = 1$. We can therefore rewrite \eqref{eq:Sigmadef} as
\bea
\Sigma(a_1, a_2) = \sqrt{1 + k\,S_k(\chi_{12})^2} - H_1\,a_1\,S_k(\chi_{12}).
\eea
Substituting the definition of $D_{12}$, \eqref{eq:angulardiameterdistance_from_chi}, we get
\bea
\Sigma(a_1, a_2) = \sqrt{1 + k\,{D_{12}}^2\,a_1^{-2}} - H_1\,D_{12}.
\eea
Finally, we replace $k$ with the parameter $\Omega_k$ defined at a reference moment $t=t_0, a= a_0$, generally taken to be the present time. Recall that $\Omega_{k,0} = - k\,{H_0}^{-2}\,{a_0}^{-2}$ [see for example \cite{Peacock_1998} Eq. (3.33)], thus
\beq
\Sigma(a_1, a_2) = \sqrt{1 - \Omega_{k,0}\,{D_{12}}^2\,{H_0}^2\, \frac{{a_0}^2}{{a_1}^{2}}}
- H_1\,D_{12}.
\eeq
Substituting the cosmological redshift $1 + z_1 = \frac{a_0}{a_1}$, we obtain \eqref{eq:SigmaFLRW}.
}

\section{DERIVATION OF FORMULA \texorpdfstring{\eqref{eq:sigma_identity}}{(\ref{eq:sigma_identity})}}\label{app:derivation}

We substitute the definitions of $\Sigma$ \eqref{eq:Sigmadef}, and the angular diameter distance \eqref{eq:angulardiameterdistance_from_chi} to the left-hand side of \eqref{eq:sigma_identity}, and after simplification get
\begin{multline}
D_{\S\LL}\,\Sigma(a_\S, a_\O) - D_{\S\O}\,\Sigma(a_\S,a_\LL) = \\ = a_\S\,\left(S_k(\chi_{\S\LL})\,C_k(\chi_{\S\O}) - S_k(\chi_{\S\O})\,C_k(\chi_{\S\LL})\right).
\end{multline}
We can use the general trigonometric/hyperbolic formula,
\beq
S_k(x - y) = S_k(x)\,C_k(y) - S_k(y)\,C_k(x),
\eeq
which holds irrespective of $k$, see \eqref{eq:Sk_def}, \eqref{eq:Ck_def}. It follows that
\beq
\textrm{LHS} = a_\S\, S_k(\chi_{\S\LL} - \chi_{\S\O}).
\eeq
The comoving distance is additive, i.e., $\chi_{\S\O} = \chi_{\S\LL} + \chi_{\LL\O}$, therefore,
\beq
\textrm{LHS} = -a_\S\, S_k(\chi_{\LL\O}).
\eeq
Making use of \eqref{eq:angulardiameterdistance_from_chi}, we obtain
\beq
\textrm{LHS} = -\frac{a_\S}{a_\LL}\,D_{\LL\O},
\eeq
which proves the \eqref{eq:sigma_identity}.

\section{DERIVATION OF THE MAXIMAL ASYMMETRY ANGLE}
\label{app:asymmetry_angle}

We calculate the difference in time of emissions according to \eqref{eq:time_of_emission_difference}, with the time delay given by \eqref{eq:time_delay}. The position of the source is given by
\begin{equation}
    \beta^{\bm{A}} = b^{\bm{A}} + \f{v_{\S}^{\bm{A}} \, t}{D_{\S\O}},
\end{equation}
and we choose $b^{\bm{A}}$ such that it is orthogonal to $v^{\bm{A}}_{\S}$ (i.e. $b$ is the impact parameter). Then, the difference in time of emissions evaluates to
\begin{equation}
\begin{split}
    \Delta t = \f{D_{\LL\O}D_{\S\O}}{D_{\LL\S}} \Big( & \f{1}{2} \big( \theta_1 - \beta \big)^2 - \Psi(\theta_1) \\
    - & \f{1}{2} \big( \theta_2 - \beta \big)^2 + \Psi(\theta_2) \Big),
\end{split}
\end{equation}
where we suppress the vector indices as the lens is axisymmetric. This can be rearranged to
\begin{equation}
\begin{split}
    \Delta t = \f{D_{\LL\O}D_{\S\O}}{D_{\LL\S}} \Big( & \f{1}{2} \big( \theta_1^2 - \theta_2^2 - 2 \beta (\theta_1 - \theta_2) \big)
    \\
    & - (\Psi(\theta_1) - \Psi(\theta_2) \Big).
\end{split}
\end{equation}
Putting in $\beta = \sqrt{b^2 + ({v_{\S}^2 \, t^2})/{D_{\S\O}}}$, $\Psi(\theta) = \theta_E^{\,\,2} \log(\theta/\theta_E)$ (point-mass lens) and $\theta$ according to \eqref{eq:theta_point_mass}, we obtain
\begin{widetext}
\begin{equation}
\begin{split}
    \Delta t = \f{D_{\LL\O}D_{\S\O}}{D_{\LL\S}} \Big[ & \f{1}{2}
    \Big( \f{1}{4} \big( \sqrt{b^2 + \f{v_\S^2 \, t^2}{D_{\S\O}^2}} + \sqrt{b^2 + \f{v_\S^2 \, t^2}{D_{\S\O}^2} + 4\theta_E^{\,\,2}} \, \big)^2
 - \f{1}{4} \big(\sqrt{b^2 + \f{v_\S^2 \, t^2}{D_{\S\O}^2}} - \sqrt{b^2 + \f{v_\S^2 \, t^2}{D_{\S\O}^2} + 4\theta_E^{\,\,2}} \, \big)^2
    \\
    & - 2 \sqrt{b^2 + \f{v_\S^2 \, t^2}{D_{\S\O}^2}} \sqrt{b^2 + \f{v_\S^2 \, t^2}{D_{\S\O}^2} + 4\theta_E^{\,\,2}} \Big) 
    -  \theta_E^{\,\,2}(\log\f{\theta_1}{\theta_E} - \log\f{\theta_1}{\theta_E}).
\end{split}
\end{equation}
\end{widetext}
This expression looks complicated, but the expansion of the difference of squares into the product of the sum and difference of the first and the second term cancels with the third term, yielding
\begin{equation}
\begin{split}
    \Delta t = \f{D_{\LL\O}D_{\S\O}}{D_{\LL\S}} \Big[ & - \f{1}{2} \sqrt{b^2 + \f{v_\S^2 \, t^2}{D_{\S\O}^2}} \sqrt{b^2 + \f{v_\S^2 \, t^2}{D_{\S\O}^2} + 4\theta_E^{\,\,2}}
    \\
    & - \theta_E^{\,\,2} \log{\f{\theta_1}{\theta_2}}
    \Big].
\end{split}
\end{equation}
This expression gives the difference in time of emissions for any position of the source. The asymmetry angle we evaluate at $t=0$ when the source is closest to the lens. The difference in time of emissions is then
\begin{equation}
    \Delta t = \f{D_{\LL\O}D_{\S\O}}{D_{\LL\S}} \Big( - \f{b}{2} \sqrt{b^2 + 4\theta_E^{\,\,2}} - \log\f{\theta_1}{\theta_2} \Big).
\end{equation}
Multiplying it by the effective angular velocity $\omega = v_{\S,\text{tr}}/(b \, D_{\S\O})$, neglecting the logarithm (as for small $b/\theta_E \ll 1$, $\theta_1 \approx \theta_2$) and disregarding the overall sign (we are interested only in the magnitude of the angle), we obtain the final result,
\begin{equation}
    \Delta \phi_{\text{max}} = \f{1}{2} \f{v}{c} \f{D_{\O\LL}}{D_{\LL\S}} \sqrt{b^2 + 4\theta_E^{\,\,2}}.
\end{equation}

\bibliography{bibliography}

\end{document}